\documentclass{article}
 \usepackage[paper=letterpaper,dvips]{geometry}	\input pack10.sty
 \usepackage{amsmath,amsfonts,graphicx}
 \usepackage{url}

 \input smallbib		

\makeatletter
\long\def\proofbox#1{\gdef\@proofbox{#1}}
\proofbox{\small{\tt\@ifundefined{sourcepath}{}{\sourcepath/}\jobname}\\%
\ifx\UNDEF\Version\else\quad v.\Version, \fi\today}

\def\proofref#1{\proofbox{\small{\tt#1\par
	[edited by Tom Toffoli for personal use]\par
	{\tt\sourcepath/\jobname}, 
	\number\month/\number\day/\number\year
	\par}}}

 \def\affil#1{\\{\small\sl#1\par}}
 \long\def\author#1{\gdef\@author{#1}}
 \author{Tommaso Toffoli ({\tt tt\char"40bu.edu})\affil{Electrical and
Computer Engineering, Boston University, MA 02215}}

 \long\def\abstract#1{\gdef\@abstract{#1}}
 \abstract{}

\long\def\@firstoftwo#1#2{#1}
\long\def\@secondoftwo#1#2{#2}
\def\@ifundefined#1{%
  \expandafter\ifx\csname#1\endcsname\relax
    \expandafter\@firstoftwo
  \else
    \expandafter\@secondoftwo
  \fi}
 \@ifundefined{leftfrac}{\def\leftfrac{.32}}{}
 \@ifundefined{ritefrac}{\def\ritefrac{.68}}{}

\def\@maketitle{\newpage\noindent\leavevmode
  \begin{minipage}[t]{\leftfrac\textwidth}
    \hrule height0pt
    \@proofbox
  \end{minipage}\hfil
 \begin{minipage}[t]{\ritefrac\textwidth}
    \hrule height0pt
    \raggedleft
    \LARGE\@title\par
    \vskip4pt
    \large\@author
  \end{minipage}
  \vskip8pt
  \ifx\@abstract\@empty\else{\vskip.5em\leftskip1.25in\parskip4pt\small\@abstract\par\vskip.5em}\fi
  \noindent
  \rule{\textwidth}{0.4pt}
  \vskip16pt}

\makeatother

\def\ifundefined#1{\expandafter\ifx\csname#1\endcsname\relax}

 \makeatletter
\def\code{\@code\frenchspacing\@vobeyspaces\verbatim@start}
\def\@code{\the\every@verbatim
 \par\noindent
  \@beginparpenalty \predisplaypenalty
  \leftskip=.4em\rightskip\z@
  \parindent\z@\parfillskip\@flushglue\parskip\z@
  \@@par
  \def\par{%
    \if@tempswa
      \leavevmode\null\@@par\penalty\interlinepenalty
    \else
      \@tempswatrue
      \ifhmode\@@par\penalty\interlinepenalty\fi
    \fi}%
  \def\@noitemerr{\@warning{No verbatim text}}%
  \obeylines
  \verbatim@font
  \let\do\@makeother \dospecials
  \everypar \expandafter{\the\everypar \unpenalty}}

 \makeatother

 \makeatletter

 \DeclareRobustCommand\em
        {\@nomath\em \ifdim \fontdimen\@ne\font >\z@
                       \upshape \else \slshape \fi}

\def\@begintheorem#1#2{\sl \trivlist \item[\hskip \labelsep{\bf #1\ #2}]}
\def\@opargbegintheorem#1#2#3{\sl \trivlist
     \item[\hskip \labelsep{\bf #1\ #2\ (#3)}]}

 \newcommand{\ie}{i.e.,}

 \mathchardef\BY="0202
 \newcommand{\by}{\ensuremath{\BY}}              

 \def\@empty{}
 \newcommand{\asin}[2][]{{
    \def\t@mp{#1}%
    \def\@cite##1##2{\marginpar{\hfil{\footnotesize$
    \ifx\t@mp\@empty\text{##2}\else\frac{\text{##1}}{\text{##2}}\fi$}\hfil}}%
\cite[#1]{#2}}}

 \def\pages#1{}

 \newcommand{\sectlabel}[1]{\label{sect:#1}}
 \newcommand{\footlabel}[1]{\label{foot:#1}}
 \newcommand{\eqlabel}[1]{\label{eq:#1}}
 \newcommand{\figlabel}[1]{\label{fig:#1}}
 \newcommand{\tablabel}[1]{\label{tab:#1}}

 \newcommand{\Chapt}[2][]{\def\t@mp{#1}%
\chapter{#2} \ifx\t@mp\@empty\else\sectlabel{#1}\fi}
 \newcommand{\Sect}[2][]{\def\t@mp{#1}%
\section{#2} \ifx\t@mp\@empty\else\sectlabel{#1}\fi}
 \newcommand{\Subsect}[2][]{\def\t@mp{#1}%
\subsection{#2} \ifx\t@mp\@empty\else\sectlabel{#1}\fi}
 \newcommand{\Foot}[2][]{\def\t@mp{#1}%
\footnote{#2} \ifx\t@mp\@empty\else\footlabel{#1}\fi}
 \newcommand{\Eq}[2][]{\def\t@mp{#1}%
\begin{equation}#2\ifx\t@mp\@empty\notag\else\eqlabel{#1}\fi\end{equation}}
 \newcommand{\Eqaligned}[2][]{\def\t@mp{#1}%
\begin{equation}\begin{aligned}#2\end{aligned}
\ifx\t@mp\@empty\notag\else\eqlabel{#1}\fi
\end{equation}}
 \newcommand{\Eqmultline}[2][]{\def\t@mp{#1}%
\begin{multline}#2\ifx\t@mp\@empty\notag\else\eqlabel{#1}\fi
\end{multline}}
 \newcommand{\Eqgathered}[2][]{\def\t@mp{#1}%
\begin{equation}\begin{gathered}#2\end{gathered}
\ifx\t@mp\@empty\notag\else\eqlabel{#1}\fi
\end{equation}}

  \newcommand{\sect}[1]{\S\ref{sect:#1}}      
 
 \newcommand{\foot}[1]{Footnote~\ref{foot:#1}}
 \newcommand{\eq}[1]{(\ref{eq:#1})}	
 \newcommand{\fig}[1]{Fig.\,\ref{fig:#1}}
 
 \newcommand{\tab}[1]{Table\,\ref{tab:#1}}

 \def\thlabel#1{\label{th:#1}}

 \def\theor#1{Theor.~\ref{th:#1}}
 \def\lemma#1{Lemma~\ref{th:#1}}

 \newcommand{\Theor}[2][]{\def\t@mp{#1}%
\begin{theorem}#2\ifx\t@mp\@empty\else\thlabel{#1}\fi\end{theorem}}
 \newcommand{\Lemma}[2][]{\def\t@mp{#1}%
\begin{lemm}#2\ifx\t@mp\@empty\else\thlabel{#1}\fi\end{lemm}}
 \newcommand{\Prop}[2][]{\def\t@mp{#1}%
\begin{proposition}#2\ifx\t@mp\@empty\else\thlabel{#1}\fi\end{proposition}}

 \long\def\endsubsection#1{\smallskip\hbox to\hsize{\leaders\hrule\hfill\ \sect{#1}}\medskip}

  \def\@arabic#1{\number #1} 

\long\def\@makecaption#1#2{
	\vskip\abovecaptionskip
	\sbox\@tempboxa{{\small #1: #2}}%
	\ifdim\wd\@tempboxa>\hsize
	    {\small #1: #2\par}
	\else
	   \global\@minipagefalse
	   \hbox to\hsize{\hfil\box\@tempboxa\hfil}
	\fi
	\vskip\belowcaptionskip}

\def\figstrut#1{\hbox to\linewidth{\vrule height#1\hfill}}

\newcommand{\Fig}[3][]{
\begin{figure}[!htb]
 \centering{\leavevmode#2}%
 \caption{#3}
 \figlabel{#1}
\end{figure}                 }

 \newcommand{\Figwide}[3][]{
 \begin{figure*}[!t]
  \centering\leavevmode#2%
  \caption{#3}
  \figlabel{#1}
 \end{figure*}                 }

\newcommand{\Tab}[3][]{
\begin{table}[!htb]
 \centering\leavevmode#2%
 \caption{#3}
 \tablabel{#1}
\end{table}                 }

\def\cstrip#1{\setbox0=\hbox{$#1$}\kern-.5\wd0\lower2pt\box0}
\def\rstrip#1{\setbox0=\hbox{$#1$}\kern-\wd0\lower2pt\box0}
\def\lstrip#1{\setbox0=\hbox{$#1$}\lower2pt\box0}
\def\tstrip#1{\setbox0=\hbox{$#1$}\kern-.5\wd0\lower\ht0\box0}
\def\bstrip#1{\setbox0=\hbox{$#1$}\kern-.5\wd0\raise\ht0\box0}
\def\Lstrip#1{\setbox0=\hbox{$\mskip2mu#1$}\lower2pt\box0}

\def\idpad{\thinspace}
\def\id{\idpad\begingroup \tt \let\do\@makeother \dospecials 
          \@ifstar{\@sid}{\@id}}
\def\@sid#1{\def\@tempa ##1#1{##1\endgroup\idpad}\@tempa}
\def\@id{\obeyspaces \frenchspacing \@sid}

\makeatother

 \input smallbib

 \def\C{{\cal C}}
 \def\Czero{{\cal C}^\circ}
 \def\Cminus{{\cal C}^-}
 \def\Cplus{{\cal C}^+}
 \def\L{{\cal L}}
 \def\del{\partial}
 \def\natural{\mathbb{N}}
 \def\integer{\mathbb{Z}}		
 \def\real{\mathbb{R}}			
 \def\pair#1#2{\langle#1,#2\rangle}

 \newtheorem{lemm}{Lemma}
 \newtheorem{theorem}{Theorem}
 
 \newtheorem{corollary}{Corollary}
 \newtheorem{conjecture}{Conjecture}
 \newtheorem{proposition}{Proposition}
 \newenvironment{proof}{\par{\sl Proof.}\quad}{\vrule height6pt width6pt depth0pt\par\medskip}

\begin{document}\date{}


 \title{When---and how---can a cellular automaton\\ be rewritten as a lattice
gas?}
  \author{$^a$Tommaso Toffoli, $^b$Silvio Capobianco, and $^c$Patrizia
Mentrasti\affil{%
 $^a$Electrical and Computer Engineering, Boston University\\
 $^b$School.\ Comp.\ Sci., Reykjavik University, Iceland\\
 $^c$Dip.\ di Matematica, Universit\`a di Roma ``La Sapienza''\\
 {\tt tt@bu.edu}, {\tt silvio@ru.is}, {\tt mentrasti@mat.uniroma1.it}}}

 \abstract{Both \emph{cellular automata} (CA) and \emph{lattice-gas automata}
(LG) provide finite algorithmic presentations for certain classes of
infinite dynamical systems studied by symbolic dynamics; it is customary to
use the term `cellular automaton' or `lattice gas' for the dynamic system
itself as well as for its presentation. The two kinds of presentation share
many traits but also display profound differences on issues ranging from
decidability to modeling convenience and physical implementability.

Following a conjecture by Toffoli and Margolus, it had been proved by Kari
(and by Durand--Lose for more than two dimensions) that any \emph{invertible}
CA can be rewritten as an LG (with a possibly much more complex ``unit
cell''). But until now it was not known whether this is possible in general
for \emph{noninvertible} CA---which comprise ``almost all'' CA and represent
the bulk of examples in theory and applications. Even circumstantial
evidence---whether in favor or against---was lacking.

Here, for noninvertible CA, (a) we prove that an LG presentation is out of
the question for the vanishingly small class of \emph{surjective} ones. We
then turn our attention to all the rest---noninvertible \emph{and}
nonsurjective---which comprise all the typical ones, including Conway's `Game
of Life'. For these (b) we prove by explicit construction that all the
one-dimensional ones are representable as LG, and (c) we present and motivate
the conjecture that this result extends to \emph{any} number of dimensions.

The tradeoff between dissipation rate and structural complexity implied by
the above results have compelling implications for the thermodynamics
of computation at a microscopic scale.}

 \maketitle

  \def\myquote#1{\hfill\begin{minipage}{3.25in}\small#1\par\end{minipage}\bigskip}
  \myquote{{\sl I do not know of any single instance where something useful
for the work on lattice gases has been borrowed from the cellular automata
field\dots.  Lattice gases differ in essence from cellular automata. A
confusion of the two fields distorts our thinking, hides the special
properties of lattice gases, and makes it harder to develop a good
intuition.}  (Michel H\'enon\cite{henon89}, with specific reference to
Wolfram\cite{wolfram86})}

\Sect[intro]{Introduction}

Cellular automata (CA) provide a quick modeling route to phenomenological
aspects of nature---especially the emergence of complex behavior in
\emph{dissipative} systems. But lattice-gas automata (LG) are unmatched as a
source of fine-grained models of fundamental aspects of physics, especially
for expressing the dynamics of \emph{conservative}\footnote
 {A dynamics is called `conservative' if it is the manifest expression of an
\emph{invertible} microscopic mechanism. It is called `dissipative' if the
underlying mechanism is not invertible to begin with, or if its invertibility
is de facto irrelevant because one is not capable or willing to maintain a
strict accounting of microscopic states---perhaps owing to lack of precise
knowledge of the initial state and the evolution laws, impredictable
influences on the part of the environment, or the sheer size of the task.}
 systems.

In the quote at the beginning of this paper, one may well sympathize with
H\'enon's annoyance: it turns out that dynamical behavior that is synthesized
with the utmost naturalness when using lattice gases as a ``programming
language'' become perversely hard to express in the cellular automata
language.  Yet, H\'enon's are visceral feelings, not argued conclusions. With
as much irritation one could retort, ``How can lattice gases differ `in
essence' from cellular automata if they are merely a subset of them? What are
these CA legacies that may `distort our thinking' and `hide the special
properties of lattice gases'? And aren't there dynamical systems that are
much more naturally and easily modeled as cellular automata?''

Today, with the benefit of twenty years' hindsight---and especially after the
results of the present paper---we are in a position to defuse the
argument. H\'enon's appeal could less belligerantly be reworded as follows:
``Even though CA and LG describe \emph{essentially the same class} of
objects, for sound pedagogical reasons it may be expedient to deal with them
in separate chapters---or even in separate books for different audiences and
applications. What is ox in the stable may well be beef on the table.''
The bottom-line message is that these two modeling approaches do not reflect
mutually exclusive strategies, but just opposite tradeoffs between the
structural complexity of a piece of computing machinery and its thermodynamic
efficiency.

\Sect[preview]{Preview}

Let $\C$ and $\L$ be the sets of dynamical systems representable,
respectively, as cellular automata and lattice gases. Our overall question
is, How are these two sets related? On one hand, we shall see that any
lattice gas can be trivially rewritten as a cellular automaton---hence
$\L\subset\C$. As for the converse issue, \ie\ \emph{how much of $\C$ is in
$\L$}, let's recall that $\C$ is naturally partitioned into three classes
which reflect fundamental categorical properties, namely,
 $$
 \begin{tabular}{@{}l|ll@{}}
	& descriptive name & categorical property\\\hline
	$\Cplus$ & invertible &	injective and surjective\\
	$\Czero$ & almost invertible & noninjective but surjective\\
	$\Cminus$ & locally lossy & noninjective, nonsurjective
 \end{tabular}
 $$
 According to a conjecture made by one of us in 1990 and proved by Kari in
1996 for the 1D and 2D cases and Durand--Lose in 2001 for the general case,
the vanishingly small class $\Cplus$ of \emph{invertible} cellular automata
is definitely ``in.'' For the other two classes, evidence has been lacking
either way\footnote
 {Except for \emph{second-order} cellular automata, already proved ``in'' by
us in 2004 (see \sect{2nd}), and a few other sporadic cases. Of course, for
any CA that was obtained from an LG as per \theor{lg2ca}, an LG presentation
is automatically available.}%
 ---which is particularly irritating for such deceptively simple CA as the
one consisting merely of a row of 2-input {\sc and} gates.

Here we first dismiss, as definitely ``out,'' the vanishingly small class
$\Czero$ of almost invertible CA---where, while the system as a whole loses
in one step a nonzero amount of information, the amount of loss \emph{per
site} is nonetheless zero.

What is left is the class $\Cminus$ of locally lossy CA; these lose in one
step a nonzero amount of information per site. This class comprises
\emph{almost all} cellular automata. Here we present a procedure for
rewriting any one-dimensional $\Cminus$ CA into an LG (which will have a more
complex unit cell, consisting of two layers and spanning more than one site
of the original CA). We also show a procedure for a simple 2D example.  Work
is in progress on an analogous construction (with a number of layers that
increases with the number of dimensions) hoped to be of general
applicability. On the basis of these partial results, we propose the
conjecture that---exception made for the paradoxical class $\Czero$ already
dismissed above---\emph{the family of dynamical systems presentable as
cellular automata and that presentable as lattice gases coincide}.

\Sect[back]{Background}

Symbolic dynamics studies a class of dynamical systems which display
topological continuity (reflecting locality of interaction) and invariance
under space and time translation (the regularity of a ``spacetime
crystal''\cite{toffoli04cryst})---and can thus be viewed as discrete,
locally-finite versions of the partial differential equations of field
theories. For example, the local map of a simple one-dimensional cellular
automaton, which is a a recurrence relation of the form
 \Eq{
	q^{t+1}_x = f(q_x^t,q_{x+1}^t)
 }
 is analogous to the well-known \emph{forward-time, forward-space}
finite-difference scheme\cite[p.~13]{strickwerda89}
 \Eq{
	q^{t+1}_x = g(q_x^t,q_{x+1}^t) = q_x^t + a\tfrac kh(q_{x+1}^t-q_x^t)
 }
 for the differential equation 
 \Eq{
	\frac{\del q}{\del t} = a \frac{\del q}{\del x}.
 }
 (One must keep in mind that in a cellular automaton the state set for
$q_x^t$ is restricted to a finite alphabet $A$, while for a differential
equation it ranges over the whole continuum $\real$.  On the other hand, the
local map $f$ of a cellular automaton is an unrestricted function of its
arguments, while the corresponding map $g$ of the finite-difference scheme
can only see its arguments through ``small differences''---$q_{x+1}-q_x$, of
the order of $h$, in space, and $q^{t+1}-q^t$, of the order of $k$, in time.)

Though symbolic dynamics systems are generally infinite, both \emph{cellular
automata} and \emph{lattice-gas automata} manage to provide \emph{finite
algorithmic presentations} for certain classes of them. These two kinds of
presentation share a number of traits but also display \emph{profound
differences}---on issues ranging from decidability to physical
implementability---to the point that one might suspect that CA and LG specify
quite different \emph{kinds} of dynamical systems (see the H\'enon quote at
the beginning of this paper).

On the other hand, for a number of reasons, some empirical and some
esthetical, we've long been entertaining the notion---the hope, if you wish,
since the evidence was extremely scant either way---that in fact CA and LG
provide just different \emph{presentations} for the \emph{same kind of
objects}. In other words, letting $\C$ denote the set of dynamical systems
representable as cellular automata and $\L$ the set of those representable as
lattice gases, one may ask, How are $\C$ and $\L$ related?  Are they
disjoint, partially overlapping, or even coincident? If the latter were true,
how come that LG presentations of even the very simplest CA are so hard to
come by?

\Fig[and-ca]{\def\XMAG{.75}\unitlength\XMAG bp%
 \def\HSIZ{318}\def\VSIZ{64}\def\HORG{-30}\def\VORG{0}%
 \begin{picture}(\HSIZ,\VSIZ)(\HORG,\VORG)
 \ifx\UNDEF\showframe\else\put(\HORG,\VORG){\framebox(\HSIZ,\VSIZ){}}\fi
 \put(0,0){\includegraphics[scale=\XMAG]{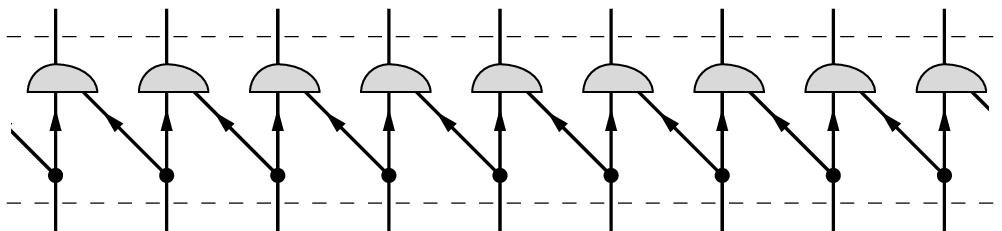}}
 \put(-16,8){\lstrip{t}}	
 \put(-28,56){\lstrip{t{+}1}}	
 \end{picture}
}
 {A simple one-dimensional \emph{cellular automaton}; this circuit can be
thought of as a \emph{presentation} of a dynamical system on the full
shift\cite{lind95}.  For concreteness, assume that the site state alphabet is
binary and the local transition function at each site, denoted by a shaded
gate, is a logic {\sc and}; the gates are fed by \emph{fanout} (or ``signal
replicator'') nodes, denoted by dots. Is there an equivalent lattice gas?}

Take, for example, the simplest nontrivial CA, namely, a string of 2-input
{\sc and} gates fed by fanout nodes (\fig{and-ca}). Until now, no one had
managed to exhibit a functionally equivalent LG or, contrariwise, prove that
such an LG cannot exist! What's perhaps even more intriguing, this very
question has never (as far as we know) been raised in the literature, even
though the analogous problem for \emph{invertible} CA had been stated and
made the object of a conjecture fifteen years ago\cite{toffoli90ica} and then
positively solved in the ensuing decade\cite{kari96,durandlose01}; and
similarly, though more recently, for \emph{second-order} CA\cite{toffoli04}.
Apparently, everyone was just as clueless about the present simple {\sc and}
example as we ourselves were until yesterday.

\bigskip

Before proceeding further it will be convenient to informally recall the
definition of `lattice gas' and the structural difference between cellular
automata and lattice gases.

\Fig[simp-lg]{\def\XMAG{.75}\unitlength\XMAG bp%
 \def\HSIZ{318}\def\VSIZ{64}\def\HORG{-30}\def\VORG{0}%
 \begin{picture}(\HSIZ,\VSIZ)(\HORG,\VORG)
 \ifx\UNDEF\showframe\else\put(\HORG,\VORG){\framebox(\HSIZ,\VSIZ){}}\fi
 \put(0,0){\includegraphics[scale=\XMAG]{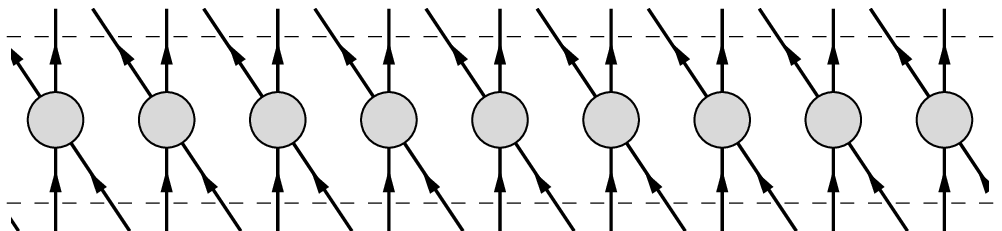}}
 \put(-16,8){\lstrip{t}}	
 \put(-28,56){\lstrip{t{+}1}}	
 \multiput(16,32)(32,0){9}{\cstrip{g}}
 \multiput(8,20)(32,0){9}{\lstrip{\scriptstyle b}}
 \multiput(24,25)(32,0){9}{\lstrip{\scriptstyle a}}
 \multiput(18,45)(32,0){9}{\lstrip{\scriptstyle b}}
 \multiput(2.5,40.5)(32,0){9}{\lstrip{\scriptstyle a}}
 \end{picture}
}
 {Format of a simple one-dimensional lattice gas. Here we may imagine $g$ to
be an arbitrary function of two binary inputs, yielding as a result two
distinct binary outputs. No fanout of signals (cf.\ the fanout nodes of
\fig{and-ca}) is permitted. The labels $a$ and $b$ distinguish the two inputs
as well as the two outputs; they actually are labels for \emph{arcs} rather
than \emph{I/O ports}.}

The simplest nontrivial\footnote
 {A CA is \emph{trivial} if it has a neighborhood of size 1, and thus, up to
a translation, the dynamics decomposes into identical independent dynamics
for each site.}
 LG have the format of \fig{simp-lg}.  Instead of having, as in a cellular
automaton (\fig{and-ca}), a single output from each node and making as many
copies of it as necessary to ``fan it out'' to the nodes that at the next
time step will use it as an input, a lattice gas \emph{does not make use of
signal fanout}; instead, each node has as many outputs (these may have
\emph{different} values, not just copies of the same one) as it has
inputs. More precisely, for a CA of state alphabet $A$ and number of
neighbors $n$ the local map is of the form $f: A^n\to A$. In an LG, on the
other hand, the state alphabet has the structure $A=A_1\by A_2\by\cdots\by
A_n$ (the factor alphabets $A_1,\dots,A_n$ need not have the same number of
elements), and the local map has the form $g: A_1\by A_2\by\cdots\by A_n\to
A_1\by A_2\by\cdots\by A_n$, with no fanout required or permitted.

\medskip

As we shall see in \theor{lg2ca}, lattice-gas automata can be thought of as a
special case of cellular automata. They were used (without being given a
special name) by \cite{toffoli77univers,toffoli77} and then extensively
investigated by Margolus and
Toffoli\cite{margolus84,toffoli87book,toffoli90ica}; one term used at that
time was `partitioning cellular automata'. In 1985, a rudimentary CA model of
fluid dynamics discovered independently by the latter research program turned
out to be quite similar to one proposed a few years before by the
hydrodynamicist Yves Pomeau and colleagues; in turn, this convergence soon
stimulated a whole industry of ``lattice-gas hydrodynamics'' research (see
\cite{toffoli90ica} for references). At that time, since the term `lattice
gas' had already been used for sundry discrete models of fluid dynamics, it
became customary to call those special cellular automata `lattice-gas
automata' or simply `lattice gases'; the term `block cellular automata' is
also in common use. The no-fanout constraint is used with the same meaning,
though in an independent line of research, in Girard's \emph{linear
logic}\cite{girard87}. There, as here, multiple uses require explicit
duplication, with all the costs (in infrastructure or running expenses) that
that may entail (see also \cite{baker94}).

\medskip

The prototypical event in a lattice gas is the collision of abstract tokens
or ``particles.'' In many practical cases this may involve a mere reshuffling
of them. A generalization of this process is when what is permuted are not
the particles but the \emph{states} of the input tuple (a reshuffling of
particle is then the special case of a permutation of the \emph{indices} of
this tuple); this ``transmutation'' may be visualized as the creation, out of
the collision of particles, of an ``excited state'' (corresponding to a
lattice-gas node), which on decaying releases particles that may differ from
the original ones.

Note that in \fig{and-ca} as well as \fig{simp-lg} the \emph{state} $q^t$ of
the system at time $t$ is the infinite collection of signals crossing the
dashed line $t$. The combinational-logic diagram drawn between the two dashed
lines constitutes a \emph{presentation} of the dynamical system, that is, a
way to indicate by an explicit, locally-finite algorithm the functional
dependency of the state at time $t+1$ on that at time $t$. Namely, for the CA of
\fig{and-ca}, using $i$ as a sequential site index, the algorithm specifies
that
 $$
	q^{t+1}_i = \text{\sc and}(q^t_i,q^t_{i+1});
 $$
 different presentations are of course possible for the same system. 
 In the LG of \fig{simp-lg}, the state-component at each site consists of a
\emph{pair} of signals, labeled $a$ and $b$, and the local algorithm
specifies that
 $$
	a^{t+1}_i = g_a(a^t_i,b^t_{i+1})\quad\text{and}\quad
	b^{t+1}_i = g_b(a^t_i,b^t_{i+1}).
 $$	
 Either in a CA or in an LG the presentation may consist of any finite number
of combinational-logic \emph{layers}, as in \fig{lg2nd}, rather than a single
layer as in the previous two figures, each layer obeying the respective
discipline for a CA or an LG.

Finally, two dynamical-system presentations are \emph{equivalent} (or
\emph{conjugate}\cite{lind95}) if the systems they describe are
isomorphic---the presentations are ``merely different views of the same
underlying object.'' Note that, in rewriting a CA as an LG, one is not
permitted to introduce supplementary variables, or `ancillae', to be used as
a scratchpad for intermediate computations (cf.\ \cite{toffoli80rc} and
\cite{bennett01}); one must make do with whatever state-variables are already
available at each site.

\medskip

To anticipate \sect{thermo}, we mention here that the above distinction
between CA and LG makes a vital difference when one turns to the physical
implementation of an algorithm. Even in ideal conditions, a fanout
node---which creates copies of a signal---needs by its very nature a source
of energy to operate; morever, as Landauer argued in \cite{landauer61},
energy is then turned into heat when signals are erased, discarded, or
destroyed---as in the {\sc and} gates of \fig{and-ca}, which take in two
binary lines but only put out one. In this sense, CA provide---for the same
functionality---\emph{cheaper mechanisms} than LGs: they are are easier to
fabricate (as we shall see in a moment), but need a power source to operate.

\bigskip

From the above definitions, one can see that

 \Theor[lg2ca]{Any LG is immediately rewritable as a CA.}

 \begin{proof}
 We use the LG of \fig{simp-lg} as a specific example, but the construction
method is fully general; starting with that LG, we gradually modify it into a
CA. Represent the pair of signals $\pair ab$ leaving or entering a node as a
single signal $q=\pair ab$ (if $a\in A$ and $b\in B$, then $q\in A\by
B$). With reference to \fig{lg2ca}, as soon a signal $q$ at site $i$ crosses
the time line $t$ we put it through a fanout element to obtain two copies of
it, and feed one copy to the node (embodying a function $g'$) at site $i$ and
the other to that at $i-1$. A node receives, through inputs
$q_i=\pair{a_i}{b_i}$ and $q_{i+1}=\pair{a_{i+1}}{b_{i+1}}$), one $a$ input
from site $i+1$ and one $b$ input from site $i$---just like a node $g$ of
\fig{simp-lg}---as well as two extra inputs, namely, one $b$ from site $i+1$
and one $a$ from $i$. We define $g'$ as the function which ignores the latter
two inputs (starred, in the figure) and otherwise acts identically to
$g$. The resulting system is a CA isomorphic to the original LG.
 \end{proof}

\Fig[lg2ca]{\def\XMAG{.75}\unitlength\XMAG bp%
 \def\HSIZ{318}\def\VSIZ{72}\def\HORG{-30}\def\VORG{0}%
 \begin{picture}(\HSIZ,\VSIZ)(\HORG,\VORG)
 \ifx\UNDEF\showframe\else\put(\HORG,\VORG){\framebox(\HSIZ,\VSIZ){}}\fi
 \put(0,0){\includegraphics[scale=\XMAG]{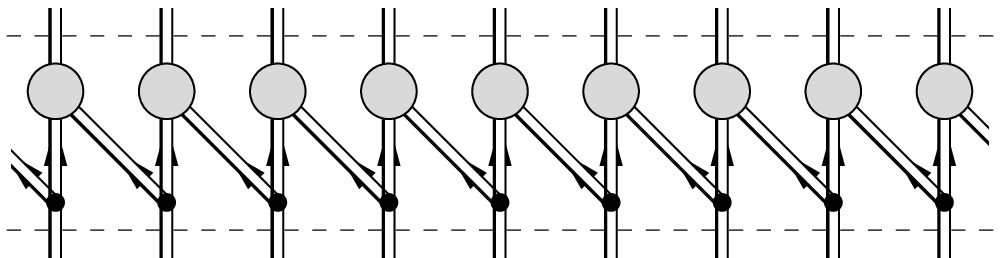}}
 \put(-16,8){\lstrip{t}}	
 \put(-28,64){\lstrip{t{+}1}}	
 \multiput(16,48)(32,0){9}{\cstrip{g'}}
 \multiput(9,40)(32,0){9}{\cstrip{\scriptscriptstyle *}}
 \multiput(26,46)(32,0){9}{\cstrip{\scriptscriptstyle *}}
 \put(134,-4){\lstrip{\scriptstyle q=\pair ab}}
 \multiput(25,30)(32,0){9}{\lstrip{\scriptstyle a}}
 \multiput(31.5,38.5)(32,0){9}{\lstrip{\scriptstyle b}}
 \multiput(7.5,34.5)(32,0){9}{\lstrip{\scriptstyle a}}
 \multiput(18.5,35)(32,0){9}{\lstrip{\scriptstyle b}}
 \end{picture}
}
 {A CA presentation of the system described by the LG of \fig{simp-lg}. Both
$a$ and $b$, bundled together as $q$, are duplicated by the fanout nodes, but
then one copy of each (starred) is ignored by $g'$, which in this way
effectively acts just as $g$.}

The outcome of this is that, despite the difference in presentation, the
functional dependence of the global state at $t+1$ on that at $t$ is the
\emph{same} in both \fig{simp-lg} and \fig{lg2ca}. Clearly, transcribing from
an LG format to a CA one is easy; how hard is the opposite direction?

\bigskip

We shall now draw a rough chart of the area we intend to explore; this area
need not extend beyond $\C$ since, by \theor{lg2ca}, all of $\L$ (whatever
its extent may turn out to be) is contained in $\C$.

 \Eq[four]{\def\XMAG{1}\unitlength\XMAG bp%
 \def\HSIZ{177}\def\VSIZ{144}\def\HORG{0}\def\VORG{0}%
 \vcenter{\hbox{\begin{picture}(\HSIZ,\VSIZ)(\HORG,\VORG)
 \ifx\UNDEF\showframe\else\put(\HORG,\VORG){\framebox(\HSIZ,\VSIZ){}}\fi
 \put(0,0){\begin{tabular}[b]{@{}c@{}|@{}c@{}|c|}
   \multicolumn{1}{@{}c@{}}{} &
     \multicolumn{1}{@{}c@{}}{\llap{inje}c\rlap{tive}} &
     \multicolumn{1}{c}{noninjective} \\\cline{2-3}
   \hbox{surjective\ }&
     \multicolumn{1}{@{\,}c@{\,}|}{$\Cplus$} &
     \vrule width0pt height 10pt$\Czero$  \\\cline{2-3}
  \begin{tabular}{@{}c@{\ }}non-\\surjective\end{tabular} &\multicolumn{1}{@{}c@{}|}{} &
  \renewcommand{\arraystretch}{2}
  \begin{tabular}{c}\\\\$\Cminus$\\\hbox to1.2in{\hfil locally lossy\hfil}\\\\\end{tabular}
  \\\cline{2-3}
 \end{tabular}}
 \put(64,114){\vector(-1,1){10}} \put(65,109){invertible}
 \put(84,95){almost invertible} \put(157,100){\vector(1,2){13}}
 \put(45,0){\includegraphics[scale=\XMAG]{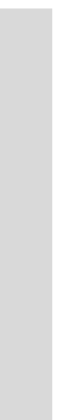}}
 \end{picture}}}
}
 In diagram \eq{four} the class $\C$ of symbolic dynamics systems presentable
as cellular automata is shown partitioned into four subclasses according to
whether or not a system is surjective (or ``onto'') and whether or not it is
injective (or ``one-to-one'').  This classification, which is of a
fundamental category-theoretical nature, dates back to the earliest attempts
to deal with cellular automata from a mathematical (rather than merely
phenomenological) viewpoint\cite{moore62,myhill63}.  The diagram's
proportions remind one that surjective systems represent a vanishing fraction
of the entire class, or a \emph{subset of measure zero}---in the sense that,
as one increases a cellular automaton's complexity in terms of size of the
state alphabet and of the neighborhood, the fraction of cellular automata
that are surjective goes to zero; similarly for injective systems.

It turns out that there are no cellular automata that are injective but not
surjective\cite{richardson72}; we indicate this by graying out the
corresponding area, which henceforth will no longer be of concern to
us. Thus, injective CA make up a vanishing subset not only of the whole set
but also of even just the surjective ones.

We shall denote by $\Cplus$ the minuscule class of systems that are
\emph{both} surjective and injective, called \emph{invertible} (or, in
physics, ``microscopically reversible'').  As we shall relate in \sect{ica},
any invertible cellular automaton can indeed be rewritten as a lattice gas;
thus all of $\Cplus$ is in $\L$.

In \sect{surj} we discuss the class $\Czero$ of \emph{almost invertible}
cellular automata---those that are surjective but not injective---and we
prove that \emph{none}  of $\Czero$ is in $\L$.

What is left is the class $\Cminus$ of \emph{locally lossy} cellular
automata---that is, those that, not being especially marked as invertible or
surjective, are (as one used to say) in ``general position'' (we'll give a
more concrete characterization of them in a moment).  These, which make up
the bulk of all cellular automata, are this paper's main concern. Only a few
are known to be presentable as lattice gases.  Here we show, by means of an
explicit construction, that all one-dimensional cellular automata of this
class are so presentable, and indicate how this construction might be
extended to multidimensional systems. On the basis of this evidence, we
conjecture that \emph{any} cellular automaton, with the exception of those of
class $\Czero$, can be rewritten as a (possibly much more complicated)
lattice gas; in other words, that $\L=\Cplus+\Cminus$ (cf.\ diagram
\eq{four}).

\medskip

In brief, concerning what parts of $\C$ are in $\L$, 
 \begin{itemize}
  \item $\Cplus$ is ``in'' (proved by \cite{kari96,durandlose01}).
  \item $\Czero$ is ``out'' (proved here).
  \item $\Cminus$ is ``in'' (proved here for one dimension; conjectured here
for the general case, with an actual working 2D example and a plausible
direction for a proof).
 \end{itemize}

\Sect[char]{Local recognition criteria for $\Cplus$ and $\Cminus$}

The categorical characterization of an invertible CA according to diagram
\eq{four}---that it be surjective and injective, that is to say, that every
global state (or configuration) must have one and only one predecessor---is
rather impractical, since configurations are countably infinite objects and
their number is uncountably infinite.

In \cite{toffoli90ica} we had provided a practical \emph{local} criterion for
recognizing invertibility---one that involved only local maps applied
to a finite number of sites; namely,

 \Lemma[toffoli/margolus]{There is an effective procedure for
deciding, for any two local maps $f$ and $f'$, whether the corresponding
global maps $F$ and $F'$ are inverses of one another.}

In a similar vein, our term ``locally lossy'' for a noninjective,
nonsurjective CA reflects the following practical local criterion
(\lemma{lossy}), which is the only one that we'll have a need for in this
paper, and which likewise involves only a CA's local map applied to a finite
number of sites.

In both cases, we feel that the reader may gain more immediate access to the
arguments and constructions discussed below by taking the local recognition
criterion as the very \emph{definition} of the kind of system in question,
and the categorical properties as an equivalent, more elegant but more
remote, characterization.

\medskip

We recall that the \emph{neighbors} of a site $i$ are those sites from which
the local map $f$ takes its arguments at time $t$ in order to compute $i$'s
new value for time $t+1$. The \emph{neighborhood}\footnote
 {Also called \emph{neighborhood function}, \emph{neighborhood index}, or
\emph{neighborhood template}.}
 $X$ is a function which, when applied to a site $i$, yields the site's
neighbors $i+j_0,\dots,i+j_h$; it can thus be thought of as the collection
$j_0,\dots,j_h$ of these offsets. (In $d$ dimensions, sites $i$ and offsets
$j$ are represented by $d$-tuples.)

We shall call \emph{patch} any collection of sites, and \emph{pattern} the
collective state of a patch.  If $I$ is a patch, $X(I)$ is defined in the
obvious way as the \emph{union} of the neighborhoods $X(i)$ for all sites $i$
of the patch. When $X$ is understood we write $\bar I$ for $X(I)$ and call it
the \emph{(causal) closure} of $I$---as it consists of all those sites the
knowledge of whose state may be needed\footnote
 {At least for \emph{some} $f$ having that neighborhood.}
 in order to determine the new state of $I$. Just as $f$ maps the state of
the neighborhood $\bar i$ of a site $i$ to that of the site itself, so does
the induced function $f^{(I)}: A^{\bar I}\to A^I$, which is but a spatial
iterate (a convolution) of $f$, map the state of a patch's $I$ closure, $\bar
I$, to the state of the patch itself. When the patch $I$ is understood, there
will be no need to distinguish between $f$ and $f^{(I)}$.  The \emph{global
map} $F$ of a cellular automaton of local map $f$ is $f^{(\integer^d)}$,
where $\integer^d$ represents the entire infinite $d$-dimensional array of
sites.
 If $p$ is a pattern on a patch $I$, then $I$ is the \emph{support} of that
pattern.
 A pattern $P$ that has no predecessors, \ie\ for which the equation
$f(P')=P$ has no solutions, is called an \emph{orphan} pattern.

\medskip

We define a cellular automaton \emph{locally lossy} if it has finite orphan
patterns, that is, if for some finite set of sites $I$ the application of $f$
to $\bar I$ fails to yield all possible states for $I$ itself.

 \Lemma[lossy]{A cellular automaton is locally lossy iff it is not surjective.}

 \begin{proof}
 If in a CA there is a finite patch $I$ that has an orphan pattern, then any
configuration containing this pattern is itself an orphan---and thus the CA
is not surjective.

To prove the converse, \ie\ that the CA is surjective if there are no finite
orphan patterns, assume that the CA is \emph{not} surjective, and thus has a
configuration $C$ with no predecessors under the global map $F$. Take a
sequence $C_0,C_1,\dots$ of finite patterns each containing the preceding
one, all agreeing with $C$ on their respective supports, and such that these
supports cover the whole space of sites. If there are no finite orphans, each
pattern $C_j$ will have, on the closure of its support, at least one
predecessor that we'll call $B_j$. If we extend in any manner $B_j$ to the
entire space we get a configuration, $K_j$. The sequence $K_0,K_1,\dots$
being defined on a compact space,\footnote
 {The configurations set of a CA with the natural product topology is
\emph{compact} by Tychonoff's theorem. Moreover, the global map of a CA
is by construction \emph{continuous} with respect to this topology.}
 has at least one accumulation point $K$, and from it one can extract a
subsequence $K_{j_0},K_{j_1},\dots$ that has $K$ as its limit. By construction,
$F(K_{j_n})$ coincides with $C$ on the support of $C_{j_n}$; by continuity,
$F(K)=C$, which contradict the \emph{absurdum} hypothesis that $C$ had no
predecessors. (This proof is patterned after one by Fiorenzi\cite[Prop.\
3.2.7]{fiorenzi00}.)
 \end{proof}

See \foot{calude} for ties to other equivalent characterizations.

\medskip

We use \emph{decidable} as a synonym for `recursive', and
\emph{semidecidable} for `recursively enumerable'. A consequence of
\lemma{toffoli/margolus} is that the class $\Cplus$ is
semidecidable. Similarly, \lemma{lossy} implies that the class $\Cminus$
as well is semidecidable. From this, it follows that the class $\Czero$
itself must be not only undecidable, but also \emph{not even
semidecidable}. In fact, if $\Czero$ were semidecidable, all three classes
$\Cplus$, $\Czero$, and $\Cminus$, which make up a partition of the universe
$C$, would be \emph{decidable}. But $\Cplus$ had been proven
\emph{undecidable} by Kari\cite{kari90}, which leads to a contradiction!


\medskip

We shall call \emph{regular} a cellular automaton for which either
invertibility or noninvertibility is eventually recognizable by local means
(cf.\ \sect{char}), and thus one of class $\Cplus$ or $\Cminus$, and
\emph{singular} one that is not recognizable in that way, and thus of class
$\Czero$. The results of this paper will make that terminology particularly
suggestive (see Proposition \ref{conj+}).

\Sect[ica]{Invertible cellular automata}

The state of the art in invertible cellular automata was reviewed in
\cite{toffoli90ica}; to avoid much repetition, we assume some familiarity
with that material and with Kari's follow-up\cite{kari05}.
 In the first of those two papers, one of us conjectured that all invertible
cellular automata can be rewritten as lattice gases. This conjecture was
subsequently proved, by Kari\cite{kari96} for two-dimensional cellular
automata and by Durand--Lose\cite{durandlose01} for the general case, in spite
of the undecidability of invertibility itself.\footnote
 {The invertibility of a dynamical system on the basis of its presentation as
a cellular automaton is undecidable---though semidecidable\cite{toffoli90ica}.}
 This undecidability nevertheless extracted a steep price; namely, there can
be no computable upper bound to the increase in complexity (in terms of size
of the state alphabet and of the neighborhood) in going from a cellular
automaton to a functionally equivalent lattice gas. What this means in
practice is that, though one can always emulate an invertible cellular
automaton by a lattice gas, the latter may need arbitrarily more complex
machinery per site.

A strictly analogous undecidability issue applies to the present quest for
lattice-gas counterparts of locally lossy cellular automata, and consequently
we must expect analogous complexity tradeoffs.

\Sect[surj]{Almost invertible cellular automata}

Here we use, for the class $\Czero$ of surjective but not invertible cellular
automata, the mnemonic \emph{almost invertible}. In fact, invertible cellular
automata (class $\Cplus$) do not lose information at all, while locally lossy
cellular automata (class $\Cminus$) lose in one step a nonzero amount of
information \emph{per site}. On the other hand, the automata of class
$\Czero$, while \emph{not} information-lossless, lose in one step such a
small amount of information over the entire \emph{infinite} array of sites
that the loss \emph{per site} is zero just as in the invertible
case.\Foot[calude]
 {This is just a more intuitive way of stating Moore and
Myhill's classic theorem\cite{moore62,myhill63}, namely, that (in Maruoka and
Kimura's terms\cite{maruoka79}), \emph{a cellular automaton's global map is
surjective if it admits of no erasable patterns}. See Calude's section
``Randomness in cellular automata''\cite[9.5]{calude94} for a review and
consolidation of these concepts.}

The simplest example is obtained by replacing the {\sc and} gates of
\fig{and-ca} with {\sc xor} gates. In that case every infinite configuration
has exactly \emph{two} predecessors, as in
 \Eq{
 \begin{gathered}
	\cdots000001111\cdots\lower4pt\hbox{$\searrow$}\\
	\cdots111110000\cdots\raise4pt\hbox{$\nearrow$}
 \end{gathered}
 \cdots000010000\cdots\\
}
 and thus the dynamics loses just \emph{one bit} of information per time step
\emph{over the entire array}.

Our question in this section is what part of $\Czero$ is in $\L$, that is,
which almost invertible cellular automata can be rewritten as lattice gases.	

\medskip

For the purposes of mathematical analysis, a chief advantage of a lattice-gas
presentation of a dynamical system over a cellular-automaton presentation is
that, with the former, the system directly inherits certain categorical
properties that can be ascertained, by mere inspection, on the presentation
itself. Namely,

 \Lemma{If a shift dynamical system admits of a lattice-gas presentation,
then it is injective (resp.\ surjective) if and only if every node of the
lattice gas is.}

 \begin{proof}
 First let us observe that if $S$ is a Cartesian product of finite sets and
$F$ is a componentwise map on $S$, that is, $S= \prod_{i\in I}S_i$ and\
$(F(s))_i = f_i(s_i)$, then $F$ is surjective iff each of the $f_i$ is
surjective. In fact, if $f_i(S_i) = S_i$ for all $i$, then $F(S) = S$; if, on
the contrary, $x_i \in S_i \setminus f_i(S_i)$ and $(s)_i=x_i$, then $s \in S
\setminus F(S)$. This is also true for injectivity: there can be two $s$'s
with the same image under $F$ if and only if there are two $s_i$'s with the
same image under $f_i$ for some $i$.

 We've seen in \sect{intro} that, in a lattice gas, the state alphabet $A$ is
the Cartesian product of $n$ factor alphabets, $A_1\by\cdots\by A_n$.  In
turn, the state set for the entire system is $=A^{\integer^d}$, where $d$ is
the number of dimensions of the site array. Using \fig{simp-lg}, where $d=1$
and $n=2$, for sake of illustration, we can write $S$ as
 \Eq{
	S = \cdots\ \by\ \underbrace{A_1\by A_2}_{x-1}\ \by\
\underbrace{A_1\by A_2}_{x}\ \by\ \underbrace{A_1\by A_2}_{x+1}\ \by\ \cdots
}
 where $x$ is the site index. Even though it is true that the local map $g$
maps $A_1\by A_2$ into $A_1\by A_2$, one may observe that, while output 1 of
site $x$ is used as input 1 of the \emph{same} site, output 2 of site $x$ is
used as the input 2 of the \emph{preceding} site, $x-1$, and so forth. In
general, the signal transport operation denoted by the arrows performs a
reshuffling of homologous state components between neighboring sites.  This
local coordinate reshuffling, from which there ensues at the global level a
mere state reshuffling within $S$, is obviously invertible, and thus
immaterial for the sake of injectivity and surjectivity. The thesis follows
from the previous observation.
 \end{proof}

Now, for a \emph{finite} set---like the state alphabet $A$ of a lattice
gas---it is evident by the pigeonhole principle that a transformation $f$ on
it cannot fail to be injective without failing to be surjective as well, and
vice versa; that is, if there is a point of $A$ that is not in the image of
$f$, then there must be some other point that has two counterimages. Thus a
lattice gas can only be either both injective and surjective---that is,
invertible---or neither---that is, locally lossy. We thus have that
$\L\subset\Cplus\cup\Cminus$, or

 \Theor
 {No almost invertible cellular automaton can be rewritten as a lattice gas.}

\Sect[typical]{Locally lossy cellular automata}

All that is left at this point is determine which locally lossy
cellular automata (these, as we've remarked, comprise \emph{virtually all}
cellular automata) can be rewritten as lattice gases. There is little in the
literature to guide one in this task.

\Subsect[2nd]{Second-order cellular automata}

Indeed the only systematic result concerns \emph{second-order} cellular
automata, where the new state of a site $x$ at time $t+1$ depend not only on
the state at time $t$ of a number of neighboring sites but also on the state
at time $t-1$ of $x$ itself, as in \fig{ca2nd}. Of course, these can be
rewritten as ordinary (\ie\ first-order) cellular automata by means of a
transformation analogous to the Legendre transform of classical mechanics:
the second-order ``Lagrangian'' system with state set $Q$ (the solid lines
traversing time $t$ in the figure) is viewed as a first-order ``Hamiltonian''
system with state set $P\by Q$ (the $Q$ component is denoted by the dotted
lines as they traverse time $t$).

\Fig[ca2nd]{\def\XMAG{.85}%
 \unitlength\XMAG bp
 \def\HSIZ{248}\def\VSIZ{176}\def\HORG{-40}\def\VORG{-30}
 \begin{picture}(\HSIZ,\VSIZ)(\HORG,\VORG)
 \ifx\UNDEF\showframe\else\put(\HORG,\VORG){\framebox(\HSIZ,\VSIZ){}}\fi
 \put(-16,-20){\includegraphics[scale=\XMAG]{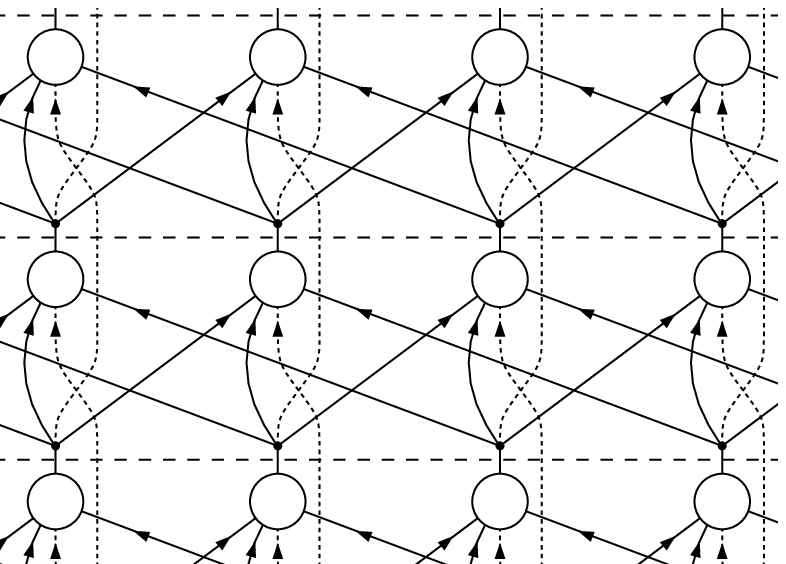}}
 \multiput(0,0)(0,64){3}{%
   \multiput(0,0)(64,0){4}{\makebox(0,0){$f$}}}
 \put(-30,138){\cstrip{t{+}1}}
 \put(-30,74){\cstrip{t}}
 \put(-30,10){\cstrip{t{-}1}}
 \put(0,-26){\cstrip{x{-}1}}
 \put(64,-26){\cstrip{x}}
 \put(128,-26){\cstrip{x{+}1}}
 \put(192,-26){\cstrip{x{+}2}}
 \end{picture}
 }
 {A second-order cellular automaton. This example is in one dimension and has
three neighbors per site (``left,'' ``center,'' and ``right of
right''), besides an input from the previous time, indicated by a dotted
line. Alternatively, the dotted signal can be thought of as a second
state-component, and the system viewed as a first-order one with two
coupled equations, like a Hamiltonian system.}

In our 2004 paper\cite{toffoli04} we showed that all second-order cellular
automata, invertible or not, can be rewritten as lattice gases. For example,
the cellular automaton of \fig{ca2nd} can be rewritten as a lattice gas
having the format of \fig{lg2nd}. There, each of the shaded blocks is a very
simple 8-input, 8-output logic function (most signals happen to go through it
unchanged), and the overall complexity of the machinery per site and time
step is not much larger than that of the original cellular automaton. Note
that each lattice-gas node straddles \emph{four} of the original
cellular-automaton sites, and it takes four layers of nodes to process one
time step. Thus, the \emph{structure} of the LG repeats itself only every
\emph{four} sites and every four layers. However, the \emph{function}
computed by the lattice gas is translation invariant for shifts of a single
site, just as the original cellular automaton. That is, we achieve a
dynamical law having a given spatial regularity in spite of using a mechanism
whose regularity pitch is four times as coarse!

\Fig[lg2nd]{\def\XMAG{.80}%
 \unitlength\XMAG bp
 \def\HSIZ{286}\def\VSIZ{180}\def\HORG{-30}\def\VORG{-10}
 \begin{picture}(\HSIZ,\VSIZ)(\HORG,\VORG)
 \ifx\UNDEF\showframe\else\put(\HORG,\VORG){\framebox(\HSIZ,\VSIZ){}}\fi
 \put(0,0){\includegraphics[scale=\XMAG]{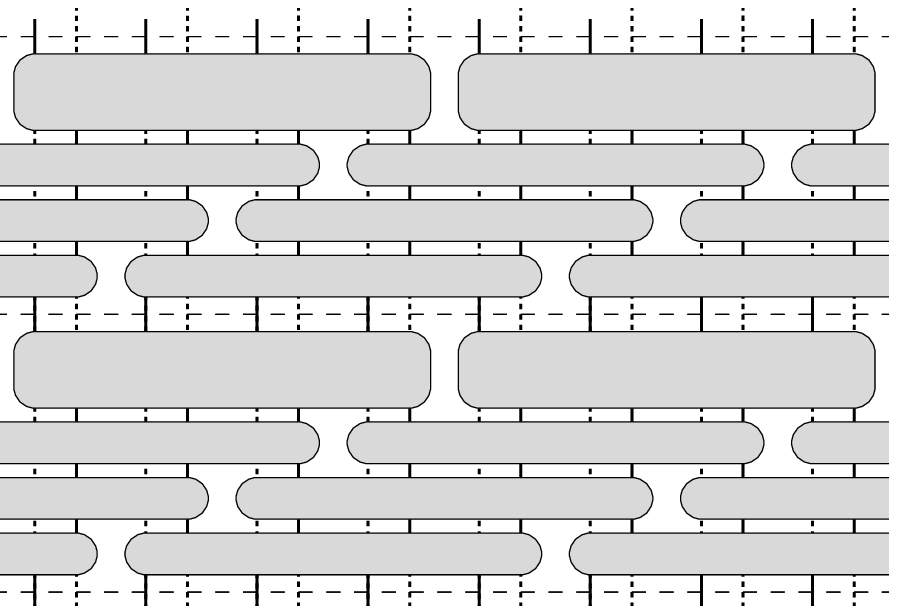}}
 \put(-18,164){\cstrip{t+1}}
 \put(-18,84){\cstrip{t}}
 \put(-18,4){\cstrip{t-1}}
 \put(16,-8){\cstrip{\cdots}}
 \put(48,-8){\cstrip{x{-}2}}
 \put(80,-8){\cstrip{x{-}1}}
 \put(112,-8){\cstrip{x}}
 \put(144,-8){\cstrip{x{+}1}}
 \put(176,-8){\cstrip{x{+}2}}
 \put(208,-8){\cstrip{\cdots}}
 \end{picture}}
 {Overall structure of a lattice gas representing a system isomorphic to that
of \fig{ca2nd}. Note that each lattice gas node straddles \emph{four}
sites, and that it takes four layers of nodes to process one time step.}

The straddling of several sites and the attendant coarsening of the
structure's translation group is a general feature of lattice-gas
presentations vis-\`a-vis cellular automata presentations---it is, as it
were, the price one has to pay to be able to sense many neighbors without
making recourse to signal fanout. In the case of second-order cellular
automata, the pitch ratio (4, in the present example) coincides with the span
of the neighborhood;\footnote
 {In more than one dimension, this ``ratio'' will be a \emph{vector} with one
component for each dimension.}
 in general, though (and, specifically, in the case of invertible cellular
automata), the pitch ratio is an unbounded function of the cellular
automaton's complexity, owing to the undecidability mentioned in
\sect{ica}. Thus one must expect to find cellular automata for which the
simplest lattice-gas counterpart, if one exists, has an arbitrarily larger
and more complex ``unit cell'' (cf.\ \cite{toffoli04cryst}) than the original
cellular automaton.  We shall see that this is also the case with locally
lossy cellular automata, and touch with hand the concrete reason why.

\Subsect[lossy]{When information loss helps}

We shall now sketch in an intuitive way the argument on which our proof and
conjecture rest. Let us go back to the ``{\sc and}'' cellular automaton of
\sect{back} and attempt to transform it into a lattice gas.  A naive
approach would be to cut it up into segments of, say, four sites each, and
turn each segment into a lattice-gas node, as in \fig{naive}.  In this way,
the neighbor links that used to connect adjoining segments are
severed:\footnote
 {By definition, a distributed dynamical system cannot have ``lateral''
transmission of information---that is, between nodes that belong to the
\emph{same} time slice---because that would introduce infinite regress
(``infinite speed of propagation of information'') and thus leave the global
state under- or over-determined.}
 neighbor data coming out of the left of each segment are lost, while the
right input to the rightmost gate of each segment remains unspecified. We may
force this input to a constant value---say, 1.

\Fig[naive]{\def\XMAG{.75}\unitlength\XMAG bp%
 \def\HSIZ{318}\def\VSIZ{80}\def\HORG{-30}\def\VORG{0}%
 \begin{picture}(\HSIZ,\VSIZ)(\HORG,\VORG)
 \ifx\UNDEF\showframe\else\put(\HORG,\VORG){\framebox(\HSIZ,\VSIZ){}}\fi
 \put(0,0){\includegraphics[scale=\XMAG]{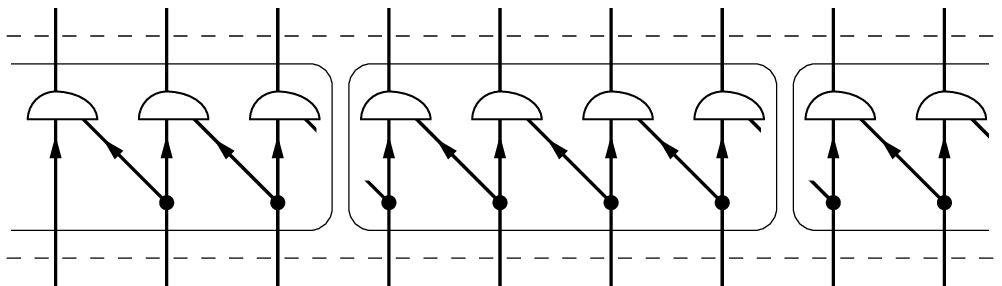}}
 \put(-16,8){\lstrip{t}}
 \put(-28,72){\lstrip{t{+}1}}
 \multiput(92,40)(128,0){2}{\cstrip{\scriptstyle1}}
 \multiput(105,34)(128,0){2}{\cstrip{\scriptstyle\by}}
 \multiput(84,69)(128,0){2}{\cstrip{\scriptstyle*}}
 \end{picture}
}
 {Here the cellular automaton of \fig{and-ca} has been sliced into four-site
segments, and each segment lumped into a lattice-gas node.}

We have now obtained a lattice gas with 4-input, \hbox{4-output} nodes each
straddling four of the cellular automaton sites. The picture inside each
node---the {\sc and} gate and the fanout junction---is merely a reminder of
what \emph{function} the node is supposed to compute, not a representation of
its internal structure. In fact a lattice-gas node, like any logic gate, has
no internal structure at all associated with it---only an assigned
correspondence (one may think of it as a lookup table) between input and
output states. Specifically, \emph{no actual fanout} (in some presumed
internal mechanism?) is implied by the use of fanout icons in the picture.

Moreover, the data transformation performed by this lattice gas from time $t$
to $t+1$ is clearly identical to that performed by the original cellular
automaton---\emph{except} for each node's rightmost output (marked with a
star), which just echos the input from the corresponding site without {\sc
and}\,ing it with its right neighbor. Our next task will be to fix this
problem without giving up the lattice-gas format.

\Fig[two-layer]{\def\XMAG{.75}\unitlength\XMAG bp%
 \def\HSIZ{318}\def\VSIZ{136}\def\HORG{-30}\def\VORG{0}%
 \begin{picture}(\HSIZ,\VSIZ)(\HORG,\VORG)
 \ifx\UNDEF\showframe\else\put(\HORG,\VORG){\framebox(\HSIZ,\VSIZ){}}\fi
 \put(0,0){\includegraphics[scale=\XMAG]{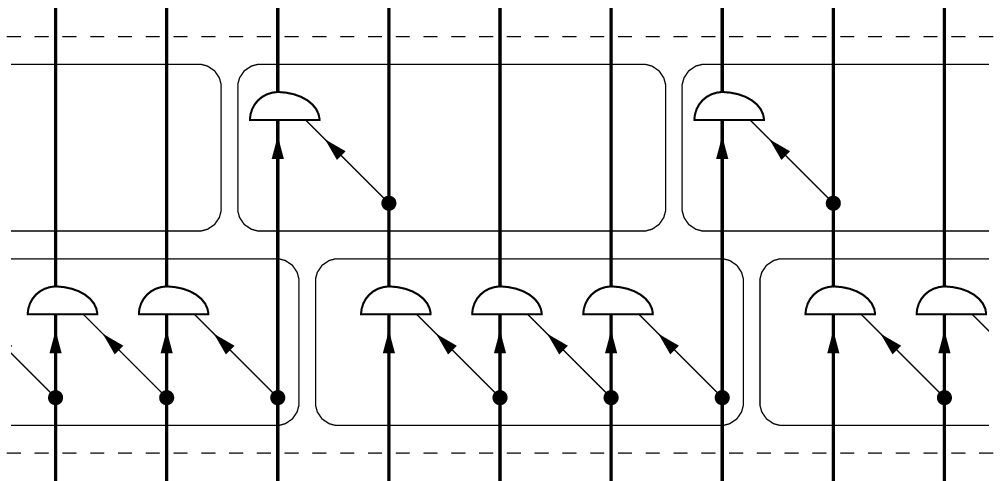}}
 \put(-16,8){\lstrip{t}}
 \put(-28,128){\lstrip{t{+}1}}
 \put(115,68){\lstrip{\scriptstyle a}}
 \put(147,68){\lstrip{\scriptstyle b}}
 \put(179,68){\lstrip{\scriptstyle c}}
 \end{picture}
}
 {Adding a second layer of nodes to bridge the gaps left from the first layer
(cf.\ \fig{naive}). Unfortunately, the lone {\sc and} gate on the second
layer gets its right-neighbor signal already as it will be at time $t+1$,
rather than as it was at time $t$.}

In order to avoid ``lateral'' transmission of information within the same
layer (which is against the rules for an LG---and indeed it would create a
path for instantaneous transmission of information across infinite distances)
we may try to add a second layer to the lattice gas, with the nodes of this
layer bridging the gaps left from the first layer, as in \fig{two-layer}.  In
this way, the lone {\sc and} gate on the second layer does get, as intended,
a right-neighbor signal as well as a center signal. However, while the
center signal is that of time $t$, as desired, the right-neighbor signal has
already been processed by the first layer and thus has the state it will
display at time $t+1$. We do get an LG, it is true, but still one with the
wrong dynamics!

\Fig[tapped]{\def\XMAG{.75}\unitlength\XMAG bp%
 \def\HSIZ{318}\def\VSIZ{156}\def\HORG{-30}\def\VORG{0}%
 \begin{picture}(\HSIZ,\VSIZ)(\HORG,\VORG)
 \ifx\UNDEF\showframe\else\put(\HORG,\VORG){\framebox(\HSIZ,\VSIZ){}}\fi
 \put(0,0){\includegraphics[scale=\XMAG]{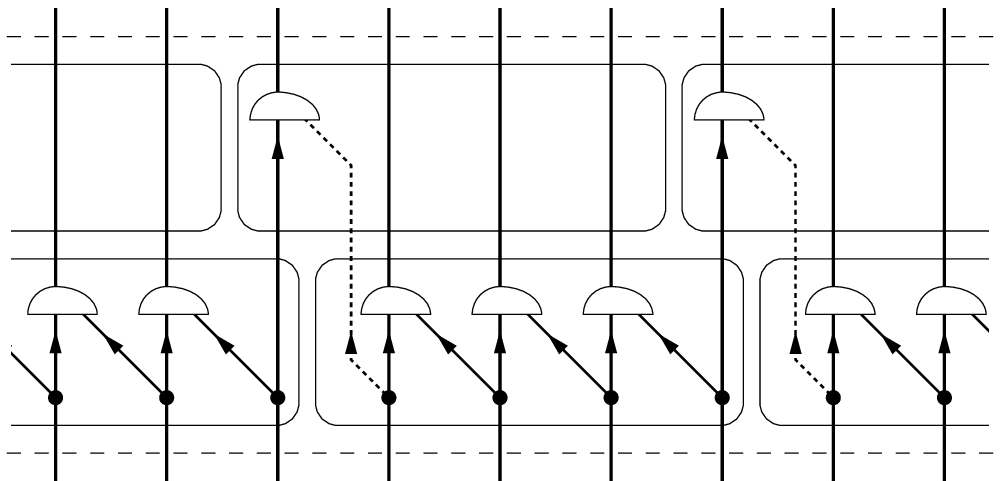}}
 \put(-16,8){\lstrip{t}}
 \put(-28,128){\lstrip{t{+}1}}
 \put(115,68){\lstrip{\scriptstyle a}}
 \put(147,68){\lstrip{\scriptstyle b}}
 \put(179,68){\lstrip{\scriptstyle c}}
 \put(115,29){\lstrip{\scriptstyle a'}}
 \put(144,138){\cstrip{\overbrace{\hbox to56pt{}}^3}}
 \end{picture}
}
 {Here we tap the state of the $a$ line at $a'$---\emph{ahead} of first-layer
processing---and forward it (dotted line) directly to the second layer.}

To obviate the above problem, we could get the right-input to the
second-layer gate by tapping line $a$ \emph{ahead} of first-layer processing,
at $a'$, so as to forward a copy of this as yet unprocessed signal (dotted
line) directly to the second layer, as in \fig{tapped}.  In this way we
certainly obtain the right dynamics--that of \fig{and-ca}---between $t$ and
$t+1$, but the intervening mechanism is no longer a lattice gas! In fact, the
first-layer nodes have four binary input lines but five output lines---and
vice versa for the second-layer nodes---while we had postulated (in
\sect{intro}) that lattice-gas nodes should have \emph{equal} input and
output state sets.  Intuitively, to perform a transformation involving four
state variables, we have taken the liberty here to introduce \emph{ex nihilo}
a fifth state variable, for intermediate storage.  From a physical viewpoint,
the dynamics depends on external degrees of freedom and is no longer
self-contained.

\medskip

Fortunately, there is one feature that hasn't yet been brought into play;
namely, that we are dealing with the class of \emph{nonsurjective} cellular
automata, which lose information on a local basis---that is, a nonzero amount
per site. Specifically, even as the input lines at time $t$ range over all
possible combinations of binary states, not all eight possible states for
lines $a,b,c$ will actually be produced; in fact, one may verify that, with
the given dynamics, state 101 will \emph{never} appear at the output. In
other words, in going from time $t$ to $t+1$ with the {\sc and} rule, the
3-bit ``channel'' $\langle a,b,c\rangle$ is not utilized at full
capacity. Can one use the spare capacity to transport some unrelated
information?  For instance, that now carried by the dotted line in
\fig{tapped}?

To this purpose, let us set up a two-layer lattice gas just as in
\fig{tapped}, but with a $k$-site wide channel, with $k$ to be chosen large
enough to provided the required capacity. As shown in \fig{and-lg}, in the
first layer a $(k{+}1)$-to-$k$ encoder $\phi$ will squeeze the extra signal
(dotted line) together with the $k$ underutilized signal lines into $k$
better utilized lines; conversely, in the second layer a $k$-to-$(k{+}1)$
decoder $\phi^{-1}$ will separate the extra signal from the other $k$ signals
and deliver it to its destination gate. Such a coder/decoder pair is termed,
in the trade, a \emph{codec}.

\Fig[and-lg]{\def\XMAG{.75}\unitlength\XMAG bp%
 \def\HSIZ{318}\def\VSIZ{200}\def\HORG{-30}\def\VORG{-14}%
 \begin{picture}(\HSIZ,\VSIZ)(\HORG,\VORG)
 \ifx\UNDEF\showframe\else\put(\HORG,\VORG){\framebox(\HSIZ,\VSIZ){}}\fi
 \put(0,0){\includegraphics[scale=\XMAG]{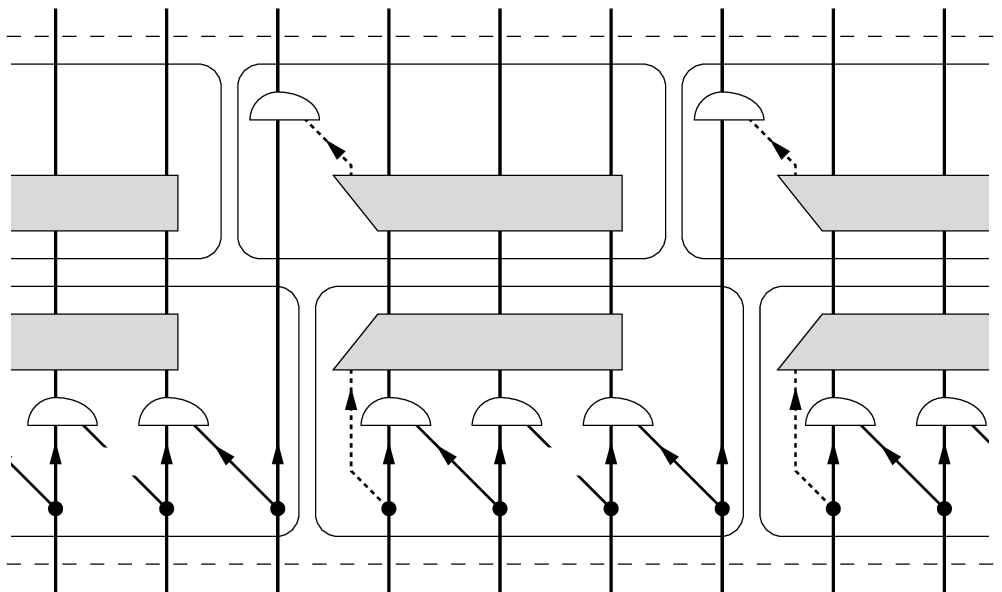}}
 \put(-16,8){\lstrip{t}}
 \put(-28,160){\lstrip{t{+}1}}
 \put(144,172){\cstrip{\overbrace{\hbox to56pt{}}^k}}
 \put(160,4){\cstrip{\underbrace{\hbox to80pt{}}_{k+1}}}
 \multiput(162,168)(-128,0){2}{\cstrip{\dots}}
 \multiput(162,95)(-128,0){2}{\cstrip{\dots}}
 \multiput(163,40)(-128,0){2}{\cstrip{\dots}}
 \multiput(162,6)(-128,0){2}{\cstrip{\dots}}
 \multiput(16, 71)(128,0){3}{\cstrip{\phi\phantom{^{-1}}}}
 \multiput(16,111)(128,0){3}{\cstrip{\phi^{-1}}}
 \put(91,58){\lstrip{\scriptstyle a'}}
 \put(115,60){\lstrip{\scriptstyle a}}
 \put(147,60){\lstrip{\scriptstyle b}}
 \put(165,61){\cstrip{\dots}}
 \put(179,60){\lstrip{\scriptstyle z}}

 \put(85,124){\lstrip{\scriptstyle a'}}
 \put(115,125){\lstrip{\scriptstyle a}}
 \put(147,125){\lstrip{\scriptstyle b}}
 \put(165,126){\cstrip{\dots}}
 \put(179,125){\lstrip{\scriptstyle z}}
 \end{picture}
}
 {To avoid running an additional signal line between first and second layer,
in the first layer we use an encoder $\phi$ to squeeze the dotted signal into
the underutilized channel provided by the next $k$ sites, and in the second
layer a decoder $\phi^{-1}$ to get this signal back out.}

What is the minimum codec width $k$ for the given {\sc and}-rule CA? For a
first rough estimate, let's observe that by taking one step of the {\sc and}
rule starting from the maximally random ensemble $U_0$ of configurations,
which has an entropy of one bit per site, we obtain an ensemble $U_1$ where
at any site a 1 will appear with a probability of 1/4. The entropy per site
$\eta$ of this new ensemble will be no more than $1/4\lg_2 4 + 3/4\lg_2 4/3$,
or $\approx .811$ (row 1 of \tab{loss})---and possibly somewhat less because
of local correlations. In fact, a tabulation of $\eta$ for blocks of
increasing length $n$ allows us to estimate an actual entropy density of
approximately .7 bits/site, and thus an entropy defect of about .3
bits/site. This is the spare ``channel capacity'' per site of the {\sc and}
CA thought of as an infinite array of binary lines filled with a noise
distribution $S_1$.

This very crude level of analysis tells us that the width of the encoder
may have to be of the order of $k=4$---since with a .3 bit/site capacity it
would take at least 4 sites to squeeze through one extra bit.  A more precise
analysis must take into consideration two additional factors:
 \begin{enumerate}
  \item We want a \emph{finite}, \emph{deterministic} encoder, which cannot
tolerate errors and yet cannot, as in Shannon's theory, fix errors by sharing
resources with adjacent blocks.  In our case, a more appropriate quantity
than the entropy $H$ of a random variable is the \emph{variety} $V$ of a
function. Intuitively, variety corresponds to the \emph{number of distinct
values} that the function takes over its domain. Like entropy, variety is
customarily expressed on a \emph{logarithmic} scale\cite{ashby56} in an
appropriate base (we shall be using base 2 for the moment). The (logarithmic)
variety $V(y)$ of the output $k$-tuple $y=\langle a,b,\dots,z\rangle$, when
the $k+1$ inputs to a block at time $t$ range over all their possible
$2^{k+1}$ values, is given by column $V$ of \tab{loss}. We see that only for
$k=8$ does the variety of the ``new state'' fall below 7 (the boldface value
6.83 on row 8), thus leaving enough space for one extra bit (for example,
$a'$) to be squeezed by the encoder into the same $k$ lines.
  \item On the other hand, $a'$ is in general to some extent correlated with
the $y$ $k$-tuple, since $a'$ is part of $y$'s ``cause.'' Thus we may expect
that $a'$ may contribute less that one bit of additional variety to the
$(k+1)$-tuple $\langle a',a,b,\dots,z\rangle$. In fact, it turns out that in
our case the variety $V'$ of the $(k+1)$-tuple $\langle
a',a,b,\dots,z\rangle$ falls to no more than $k$ already for $k=4$ (boldface
value 4 for $V'$ on row 4). That is the threshold we were looking for!
 \end{enumerate}

 \Tab[loss]{
\small
 $\begin{array}{r|rrr|r}
  k &\multicolumn{1}{c}{H}&\multicolumn{1}{c}{\eta}&\multicolumn{1}{c}{V}&\multicolumn{1}{c}{V'}\\\hline
  1&  0.81& .811&	 1.00&  1.58|\phantom{0}2\\
  2&  1.55& .774&	 2.00&  2.32|\phantom{0}3\\
  3&  2.25& .750&	 2.81&  3.17|\phantom{0}4\\
  4&  2.95& .737&	 3.58&{\bf4.00}|\phantom{0}5\\
  5&  3.65& .730&	 4.39&  4.81|\phantom{0}6\\
  6&  4.35& .725&	 5.21&  5.61|\phantom{0}7\\
  7&  5.05& .721&	 6.02&  6.43|\phantom{0}8\\
  8&  5.75& .718&     \bf6.83&  7.24|\phantom{0}9\\
  9&  6.45& .716&	 7.64&  8.05|10\\
 10&  7.15& .715&	 8.46&  8.86|11\\
 11&  7.84& .713&	 9.27&  9.67|12\\
 12&  8.54& .712&	10.08& 10.48|13\\
 13&  9.24& .711&	10.89& 11.30|14\\
 14&  9.94& .710&	11.70& 12.11|15\\
 15& 10.64& .709&	12.51& 12.92|16\\
 16& 11.34& .709&	13.32& 13.73|17
\end{array}$
 }
 {Total entropy $H$, entropy density (per site) $\eta$, and variety per site
$V$ of the output pattern of width $k$ produced in one step by the ``{\sc
and}'' cellular automaton of \fig{and-ca} from a maximally-random input
neighborhood of width $k+1$. The last column, $V'$, gives the
variety when the output $k$-tuple is augmented by the input signal $a'$
(see \fig{and-lg}). Note that, because of internal correlations, this extra
binary line only adds about 1/2 bit of variety.}

We conclude that, for the CA in question, an equivalent LG presentation with
two layers of blocks of width $n=k+1$ and an appropriate \emph{codec} of
width $k$ is possible as soon as $k\geq4$.

\Sect[1D]{The one-dimensional case in general}

The foregoing construction was given for a simplest nontrivial CA (cf.\
\sect{back}), which
 (a) is one-dimensional,
 (b) uses a 2-state alphabet (the Boolean values 0 and 1),
 (c) has a 2-element neighborhood template (consisting of the two offsets $0$
and $1$), and
 (d) uses the {\sc and} function as a local map on this neighborhood.

Here, while remaining within one dimension, we shall generalize that
construction to any \emph{state alphabet} $A$, \emph{neighborhood} $X$, and
\emph{local map} $f$. In subsequent sections we shall pursue the most general
goal along these lines, \ie\ a construction method that applies to any
\emph{number of dimensions}.

In one dimension, the \emph{diameter} of the neighborhood is the distance
between leftmost and rightmost neighbors, or $j_h-j_0$.  (In the {\sc and} CA
used in \sect{lossy} the offsets were 0 and 1 and thus the neighborhood
diameter was $h=1-0=1$.) A neighborhood may be \emph{sparse}, in the sense
that some offsets between the lowest and the highest are missing; the
diameter is not affected by sparseness. Henceforward, without loss of
generality, we shall assume that the neighborhood is not sparse (gaps may
always be filled by dummy neighbors), and shall add a constant to all offsets
so as to make the least offset 0. With this convention, the neighborhood
diameter coincides with the highest offset $h$, where $h+1$ is the number of
neighbors.

\medskip

The construction template we propose below for the general one-dimensional
case, illustrated by the sequence of steps of \fig{k+h}, has two adjustable
parameters. The first is the \emph{neighborhood diameter} $h$; the second,
the \emph{codec size} $k$ (\ie\ the number of sites spanned by it). The
corresponding block size will be $n=k+h$.

\Fig[k+h]{\def\XMAG{.63}\unitlength\XMAG bp%
 \def\HSIZ{400}\def\VSIZ{392}\def\HORG{-16}\def\VORG{-28}%
 \begin{picture}(\HSIZ,\VSIZ)(\HORG,\VORG)
 \ifx\UNDEF\showframe\else\put(\HORG,\VORG){\framebox(\HSIZ,\VSIZ){}}\fi
 \put(0,0){\includegraphics[scale=\XMAG]{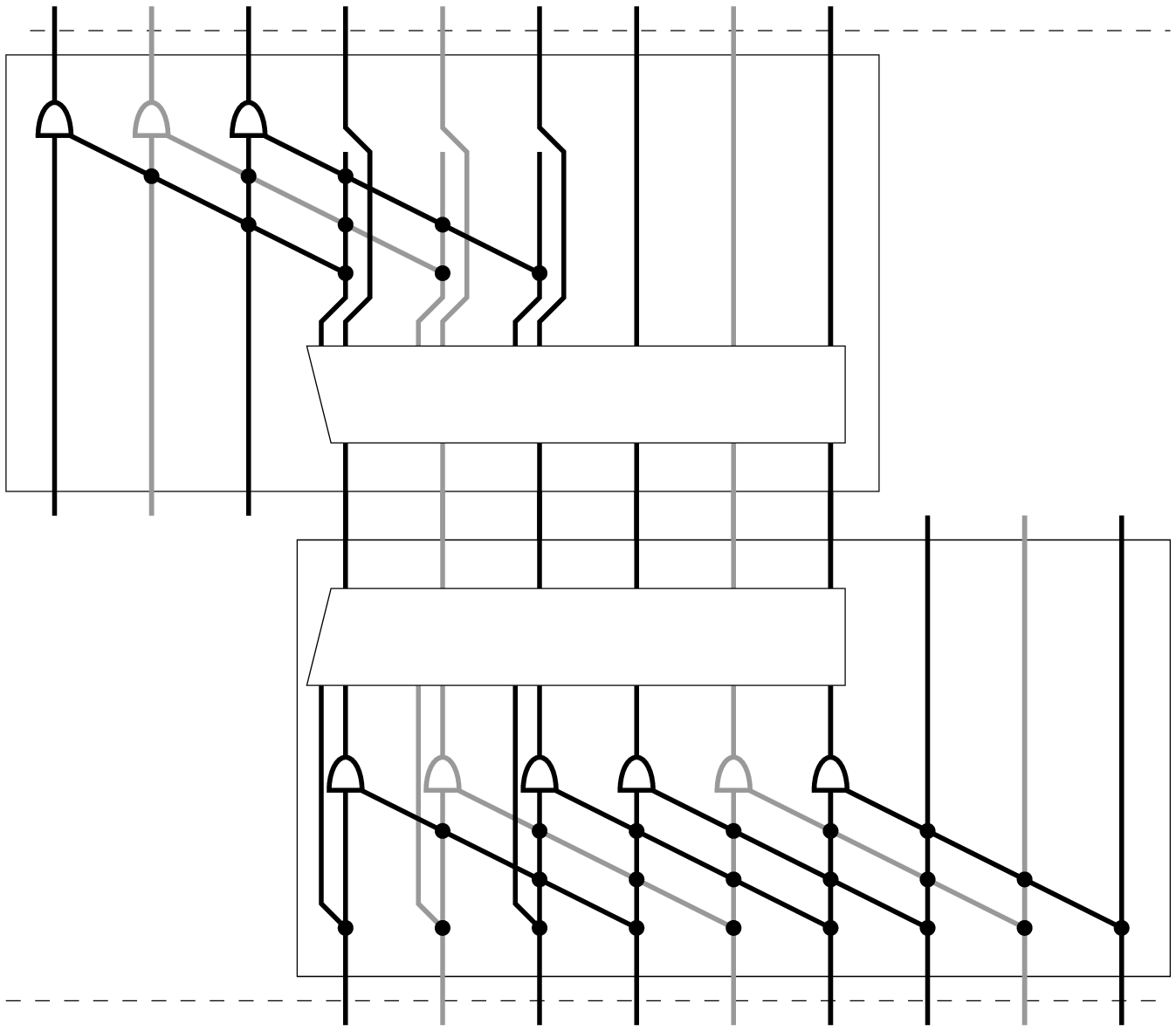}}
 \put(-12,8){\lstrip{t}}
 \put(-20,328){\lstrip{t{+}1}}
 \put(144,348){\cstrip{\overbrace {\hbox to 44pt{}}^h}}
 \put(240,348){\cstrip{\overbrace {\hbox to 44pt{}}^{k-h}}}
 \put( 48,338){\cstrip{\overbrace {\hbox to 44pt{}}^h}}
 \put(192,338){\cstrip{\overbrace {\hbox to105pt{}}^k}}
 \put(336,  6){\cstrip{\underbrace{\hbox to 44pt{}}_h}}
 \put(192,  6){\cstrip{\underbrace{\hbox to105pt{}}_k}}
 \put(144,-4){\cstrip{\underbrace{\hbox to 44pt{}}_h}}
 \put(240,-4){\cstrip{\underbrace{\hbox to 44pt{}}_{k-h}}}
 \put(180,128){\lstrip{\phi}}
 \put(180,208){\lstrip{\phi^{-1}}}
 \end{picture}
}
 {General template for converting into a two-layer lattice gas an arbitrary
one-dimensional locally lossy cellular automaton. The dome-like ``gate''
denotes the local map $f$. The parameter $h$ is the neighborhood diameter;
$k$ is the size of the codec, and $n=k+h$ the block size. Note how, in the
second layer, neighbor inputs other than the 0 neighbor are not taken from
the new-state lines but from the corresponding old-state lines, which had
been store-and-forwarded from the first layer. Notation conventions are
explained below.}

In \fig{k+h} we have used two conventions to avoid clutter.  First of all,
instead of having $h+1$ neighbor wires converge onto a gate as inputs (as,
for instance, in \fig{ca2nd}), we have retained a separate wire only for the
neighbor of offset 0; all other $h$ neighbors are picked up by a side
extension of the gate itself which senses them but also lets them go through,
so that they may be sensed in turn by all the gates of which they are
neighbors. We thus have a notation suggestive of a cross-point matrix
connecting vertical ``source'' wires to horizontal ``destination''
wires. This notation is standard in reversible logic\cite{toffoli80rc} and
quantum computing\cite{barenco95}.
 Secondly, instead of indicating an indeterminate number of repetitions of a
certain feature (such as the circuitry of one site) by explicitly giving the
first and the last occurrence of the feature and implying all the
intermediate occurrences by ellipsis dots (``\dots''), as in \fig{and-lg}, we
imply the intermediate occurrences by a \emph{single} grayed-out icon, as well
as a brace indicating how many occurrences are intended in all. Thus, the $h$
neighbors of a site (not including the site itself, or neighbor 0) are
indicated by an icon for neighbor 1, one for the last neighbor $h$, and a
single grayed-out icon standing for the remaining $h-2$ neighbors.

\bigskip For the construction to work as intended, the $k$ lines running from
the encoder $\phi$ in the first layer to the decoder $\phi^{-1}$ in the
second must be able to carry, distributed among themselves, not only the
corresponding $k$ new states just computed within the first layer, but also
information about the \emph{old} state of the leftmost $h$ of those $k$
sites, which is still needed for updating $h$ left-over sites in the second
layer. The encoder must be wide enough for this purpose. To show that a $k$
large enough to provide the required extra capacity can always be found we
shall be making use of \emph{Fekete's lemma}\cite{fekete23}, which we restate
here in a form similar to that used by van Lint and Wilson\cite{vanlint92}.

 \Lemma[fekete]{{\rm (Fekete)}\quad Let $g:\natural\to\real^+$ be a function
for which $g(i+j)\leq g(i)g(j)$ for all $i,j\in\natural$. Then
$\lim_{i\to\infty}g(i)^{1/i}$ exists and equals $\inf_{i\geq1}g(i)^{1/i}$.}

 \Theor[1D]
 {Any one-dimensional locally lossy cellular automaton can be rewritten
as an isomorphic lattice gas.}

 \def \Ibar{\bar I}

 \begin{proof} Consider a one-dimensional CA of state alphabet $A$ of
cardinality $a$, neighborhood $X$ of diameter $h$, and local map $f: A^X\to
A$.

Given a patch $I$ and its closure $\bar I$, for any state $x\in A^{\Ibar}$
the local map $f$ will determine the corresponding new state $y=f(x)$ (with
$y\in A^I$). Observe that, if the CA is not trivial, at least one state of
$I$ must have more than one pre-image among the states of $\Ibar$. On the
other hand, there may be states of $I$ that do not arise from \emph{any}
state of $\Ibar$; these are $I$'s orphans. By definition, a CA is locally
lossy if for some $i$ the corresponding patch $I$ has orphans.

Let us denote by $g(i)$ the variety of the range of $f$, that is, the number
of distinct states for $I$ that actually arise from $y=f(x)$ as $x$ runs over
$f$'s domain, $A^{\Ibar}$. Consider now two adjacent patches, $I$ and $J$,
spanning, respectively, $i$ and $j$ sites, and their juxtaposition $IJ$ of
length $i+j$. If the $I$ portion of a pattern on $IJ$ happens to be an orphan
of $I$, then \emph{that entire pattern} is an orphan for $IJ$, immaterially
of whether or not the $J$ portion of it is an orphan for $J$. It immediately
follows that $g(i+j)$ can be \emph{at most} as large as $g(i)g(j)$, and
Fekete's lemma as given above applies verbatim.

If we now express variety in a logarithmic fashion (see \sect{lossy}) using
$a$ for a base---that is, if variety is represented by the quantity
$v(i)=\log_a g(i)$---then the appropriate form of Fekete's lemma to be
applied to this quantity is the sub-additive (rather than sub-multiplicative)
one, that is,
 $$
	v(i+j)\leq v(i)+v(j)\ \ \Longrightarrow\ \ 
	\exists_{\eta\in\real}\lim_{i\to\infty}v(i)/i=\inf_{i>1}v(i)=\eta.	
 $$
 Since the variety $v(i)$ is in any case comprised between $i$ (when the
range of $f$ is all of $a^i$) and 0 (when all of $I$'s neighborhood states
map into one and the same state for $I$), the above limit $\eta$ is a number
between 0 and $1$.

For the construction of \fig{k+h} to work, the $k$ lines of the codec (which
collectively have a variety capacity of $k$) must be able to distinguish
between all effectively occurring combinations of values for the new state of
$k$ sites together with the old state of the adjacent $h$ neighboring
sites. Therefore, the joint variety $\tilde k$ of all these combinations
must not exceed the variety capacity $k$ of the codec ``channel.'' An upper
bound to $\tilde k$ is given by the sum of the marginal varieties, namely,
$h$ for the $h$ neighbors and $k'$ (to be determined below) for the new
state of the  $k$ sites.

If the CA is locally lossy, then $v(i)$ is eventually less than $i$, and thus
$\eta<1$.  In this case, given any desired variety $h$, and $\epsilon>0$ such
that $\eta'=\eta+\epsilon<1$, for large enough $k$ we have $v(k)/k<\eta'$ and
$k-v(k)>k(1-\eta')\geq h$. In other words, for any neighborhood diameter $h$
we can always find a patch $K$ of length $k$ having a \emph{variety defect}
$k-v(k)$ at least as large as $h$. A codec built on this patch will be able
to transmit through $k$ lines the variety of the new state of $k$ sites
together with that of the old state of $h$ of these sites.
 \end{proof}

We can apply \theor{1D} to Wolfram's so-called ``rule 110''\cite{wolfram02}
and conclude that
 \begin{corollary}
 There exist noninvertible computation-universal cellular automata that can
be isomorphically rewritten as lattice gases.
 \end{corollary}
 \begin{proof}It is easy to verify that a codec of sufficient capacity for
the construction of \fig{k+h} to work---with $h=2$ and rule 110 as a local
map---can be found as soon as $k\geq17$.
 \end{proof}

\Sect[2D]{A two-dimensional example}

\Figwide[flipbook]{
 \def\Step#1#2{#2{\small#1} \raise3pt\hbox{$\vcenter{%
    \hbox{\includegraphics[bb=64 118 384 406,scale=.33]{fig/flip#1.eps}}}$}}
 \def\step#1{\Step{#1}{\relax}}
 \def\stpA#1{\Step{#1}{\begin{picture}(0,0)\put(-6.5,60){\lstrip{t{=}0- - - - - - - - - - -\null}}\end{picture}}}
 \def\stpZ#1{\Step{#1}{\begin{picture}(0,0)\put(-6.5,-43){\lstrip{t{=}1- - - - - - - - - - -\null}}\end{picture}}}
 \def\insert{\framebox{\parbox{3.52in}{The new state of a site, indicated by a
circle, is a function of the current state of its four neighbors,
collectively indicated by four dots, according to the neighborhood
template \lower3pt\hbox{\includegraphics[scale=.33]{fig/neigh.eps}}\ .}}}

 \def\arraystretch{7.25} \tabcolsep5pt
 \begin{tabular}{@{}c|c|c|c@{}}
  \stpA{1b} & \step{2b} & \step{3b} & \step{4b} \\
  \step{1c} & \step{2c} & \step{3c} & \step{4c} \\
  \step{1d} & \step{2d} & \step{3d} & \step{4d} \\
  \step{1e} & \step{2e} & \step{3e} & \step{4e} \\\hline
  \step{2a} & \step{3a} & \step{4a} & \step{5a} \\
  \multicolumn{3}{c}{\insert}       & \stpZ{5b}

 \end{tabular}
}
 {Construction of a 5-layer lattice gas for the 2\by2 {\sc and} cellular
automaton.}

Before attempting to extend the above result to an arbitrary number of
dimensions, it will be expedient to verify at the very least that the
one-dimensional construction of \sect{lossy}, for a 2-input {\sc and}
cellular automaton, can be leveraged (with some creativity) to the analogous
two-dimensional case, that is, a (2\by2)-input {\sc and} automaton.

\bigskip

In the 1D case of \sect{lossy}, our CA-to-LG construction yielded a 2-layer
lattice gas that employed one codec to negotiate the information transfer
between layers 1 and 2.  The codec was four sites wide, and in either layer
the block spanned by the LG node was five site wide---the 1 extra site
corresponding to the neighborhood diameter, which was, in that case, 1.

In the present 2D case, five layers (instead of merely two) will be required
for the state of the lattice gas to advance through one time step of the
cellular automaton.  Four different codecs will be needed to negotiate the
information transfer between adjacent layers, \ie\ $1\to2$, $2\to3$, $3\to4$
and $4\to5$. As for the size of the LG blocks, which is determined by the
size of the smallest codecs that will do the job, let us remark that the most
critical passage is from layer 1 to layer 2 (as we shall see, much less
``spare capacity'' is needed at the next three passages).  Using the lossiest
(the ``most noninvertible'') nontrivial CA rule on a 2\by2 neighborhood, namely
the {\sc and} of the four neighbors,\footnote
 {This rule maps the 16 neighborhood states into 15 zeros and 1 one, while
any surjective rule (and thus any invertible one) would need to map the 16
states into 8 zeros and 8 ones---see the ``balanced map'' criterion in
\cite{maruoka79}.}
 a codec of size 4\by4 will have enough capacity to pass the required
edge-neighbor information from layer 1 to layer 2.\footnote
 {With 16 binary lines, this 4\by4 codec has a capacity for $2^{16} = 65,536$
different messages, which is more than enough for the 52,886 different
messages (combinations of new and old site values, as explained in point 2
near the end of \sect{lossy}) which we need to send through the codec from
layer 1 to layer 2.}
 Therefore, since the local map is of the format $2\by2\to1\by1$, and thus
has a neighborhood diameter of 1 in either dimension, to host a codec of size
4\by4 we shall need an LG block size of 5\by5 (cf.\ the analog situation in
the 1D case, as in \fig{and-lg}).

In \fig{flipbook} we describe the function computed by each layer in terms of
a sequence of logical steps (4 for the first layer, 2 for the last, and 5 for
each of the intervening three).  We stress again that such a sequence gives a
\emph{heuristics} for identifying that function but is not meant to be a
description of an \emph{internal mechanism} for it.  Each layer will consist
of LG nodes each simply taking its inputs from a block of 5\by5 sites and
returning its 5\by5 outputs to a block of the same size; there is no
\emph{logical} need to analyze this function as a ``computation,'' \ie\ as a
series/parallel composition of simpler functions. What's more, in a
\emph{physical} implementation of the function, in order to have the maximum
thermodynamic efficiency (see \sect{thermo}) it may be essential to make use
of a custom physical interaction that makes the 5\by5 signals interact with
one another \emph{all at once}! Whatever the heuristics, in the end it may be
more appropriate to visualize each node merely as an
indivisible interaction---a mere lookup table.

\def\layer#1#2{\medskip\noindent{\bf Layer #1} (#2)\par\smallskip}

\layer1{b,c,d}

In \fig{flipbook}, which may be viewed literally as a ``flipbook,'' we enter
layer 1 immediately after $t=0$; a dot indicates that a site is in the
``old'' ($t=0$) state. At step 1b the sites are grouped into 5\by5 blocks. At
1c, we compute the ``new'' state (the one that will be appropriate for
$t=1$), indicated by a circle, of all the sites that can view their entire
2\by2 neighborhood \emph{within that block}; a dot within a circle means that
we are at present in possession of the old as well as the new state for a
given site.

Next (1d) we ask ourselves which of the old states will still be needed for
eventually computing the new state of the sites which we haven't been able
to update yet; the 3\by3 old states which have exhausted their function may
now be dropped from consideration. At 1e we recognize that, owing to the
local noninvertibility of the {\sc and} rule, the variety of the array of
4\by4=16 new states is less than 16 bits. Therefore, the 16 lines
corresponding to those sites have enough ``channel capacity'' to carry some
extra information, specifically, the 7 old states which we are still carrying
over. We thus construct, in analogy with what we did in \fig{and-lg}, an
encoder (gray area) that compresses those 16+7 binary lines into just 16 lines
which are passed on to the next layer. The 9 not-yet-updated sites that make
the unshaded border are passed on verbatim to the next stage.

\layer2{a,b,c,d,e}

We now shift the block boundaries one site leftwards. At 2a the 16 encoded
lines are picked up by a matching decoder which expands them into the
original 16+7 lines (step 2b). At 2c we update as many new additional sites
as possible on the basis of old states that have just been made accessible
within the block by shifting its boundaries; and at 2d we discard the old
states for which we have no further need. Finally (step 2e) we prepare an
encoder for squeezing through to the next level the 5 old states that still
accompany their new states. Note that now we have more channel capacity
available (deriving from a 5\by4 rather than a 4\by4 array of new states) for
fewer lines to squeeze through (5 instead of 7); in other words, things
become easier once cleared the information bottleneck of the first layer.

\layer3{a,b,c,d,e}

Block boundaries are now shifted one site upwards. Layer 3 proceeds much like
layer 2, with just one novelty. At the moment (step 3e) of packing 20 new
states plus 2 old states into a 4\by5 encoder, we realize that one site
outside of the nominal span of the encoder has both old and new state with it,
while its one line, to be carried over verbatim to the next layer, will have no
room for both. What we do then is let the \emph{old} state of that site go
through in place, but ``tuck away'' into the encoder the \emph{new} state, as
indicated by the dashed line.

\layer4{a,b,c,d,e}

Block boundaries are shifted one site leftwards. Everything proceeds like
in layers 3 and 4.

\layer5{a,b}

We shift block boundaries rightwards, \ie\ back were they were at the end of
layer 3. We unpack everything (step 5a) from the decoder, including the new
state that had been tucked away at step 3e, and at 5b we return this state to
its original site. Now ($t=1$) that there are no more old states left we are
ready to rechristen the new states `old states' and start a new 5-layer
cycle.

\bigskip

As long as the neighborhood remains 2\by2, it is clear that the above
construction pattern will work for a larger state alphabet and an arbitrary
local map,\footnote
 {Though, of course, still one that yields a locally lossy CA.}
 just as in \sect{1D}; all that will be required is a large enough $k\by k$
block size.

\Sect[conj]{General case}

The 2D construction of \sect{2D} can apparently be generalized to an
arbitrary neighborhood size; that would make our 2D result as general as the
1D one. One may then wonder whether an analogous result holds for \emph{any}
number of dimensions. In support of that thesis, we'd like to observe that,
as long as one retains the construction strategy implied by \sect{1D} and
\sect{2D},
 \begin{enumerate}
 \item The strategy does not depend on the size $a$ of the state alphabet;
the latter only influences the parameter $k$---the size of the codec.
 \item The strategy does not depend on the diameter $h$ of the neighborhood;
the latter only determines the thickness of the ``boundary layer'' of
neighbors whose old state has to be carried through until no longer needed.
 \item The strategy does not depend on the specific local map $f$; the
latter, assumed to be locally lossy, only influences, through the extent of
the information losses (gaged by the Fekete's lemma's limit $\eta$ of
\theor{1D} or its multidimensional counterpart as in \lemma{multi}), the size
of the codec.
 \item The existence of codecs of appropriate sizes for the different stages
is guaranteed by the multivariate version of Fekete's
lemma (below).
 \end{enumerate}

 \Lemma[multi]{{\rm(multivariate Fekete's lemma;
Capobianco\cite{capobianco07})}\quad Let
\nolinebreak\hbox{$f:\natural^d\to\real^+$} be a function of $d$ variables,
subadditive in each, \ie\ for any $j$

\noindent{\small \let\olddots\dots \def\dots{{\scriptstyle\olddots}}%
$f(x_1,\dots,x_j+y_j,\dots,x_d)\leq f(x_1,\dots,x_j,\dots,x_d)+f(x_1,\dots,y_j,\dots,x_d)$.}

\noindent Let $\eta=\inf_{x_1,\dots,x_d\geq1} f(x_1,\dots,x_d)/x_1\cdots x_d$.
For any $\epsilon>0$ there exist $k_1,\dots,k_d$ such that, if $x_j>k_j$ for
all $j$, then $f(x_1,\dots,x_d)/x_1\cdots x_d<\eta+\epsilon$.}

Therefore, each of the stages, analogous to those of \sect{2D}, that may be
needed for the $d$-dimensional case, is feasible for a large enough block
size $(n_1,n_2,\dots,n_d)$. All that is missing at this point (but this may
well turn out to be a tall order) is the guarantee that, for any $d$, an
appropriate finite sequence of ``progressive partial updating'' layers may be
constructed so as to ultimately achieve, in a finite number of steps, a
complete updating of the CA.

\medskip

A bit of common sense will remind one that
 \begin{itemize}
 \item Just by looking at the increase in logical and graphical construction
complexity in going from 1D to 2D, one may want an automated
\emph{proof}---or even an automated \emph{proof generator}---to tackle the
complexity of the sequence of steps needed for more dimensions and organize
it in a comprehensible way. For instance, (a) Will the number of layers be
essentially proportional to the number of dimensions, and thus to the number
of (hyper)faces of a block? (b) Will one also need extra layers to account
for (hyper)edges, \dots, and so forth down to the 0-dimensional vertices?
 \item Numerical verification of representative test cases is invaluable in
complicated proofs. (``Pure'' mathematicians may argue otherwise, but we
suspect that, far from shunning such verifications, they actually do them
automatically and subconsciously.)  However, the computational burden of such
tests grows exponentially (and \emph{extremely} fast if one takes block
\emph{side} as a parameter), to the point that, even for Conway's ``Game of
Life''\cite{gardner70}---a moderately lossy 2-state, 3\by3-neighborhood
cellular automaton---the effective construction of the codecs is a well-nigh
intractable task!
 \item There are encouraging and discouraging precedents in this area.  As we
already mentioned at the beginning of \sect{preview}, a related conjecture,
which had taken six years to prove for one and two dimensions, took another
six years and a very complex construction to prove in the general case.  On
the other hand, in another argument entailing, again, surjectivity and
injectivity in cellular automata, the authors of \cite{amoroso72} had thought
that their techniques were ``in principle extendable to arrays of higher
dimension;'' since, however, these techniques were ``difficult to manage
beyond dimension one,'' they expected that ``generalizations of their results
to higher dimensions'' would ``most likely require a different approach.'' In
the end (that is, almost twenty years later), the conjecture turned out to be
wrong and its object undecidable\cite{kari90}!
 \end{itemize}

The evidence given, together with more work in progress on actual
multi-dimensional exhibits, is strongly suggestive but not complete. For the
moment, we prefer to propose the general case only as a conjecture, namely,

 \begin{conjecture}\label{conj}
 Any locally lossy cellular automaton can be rewritten as an isomorphic
lattice gas.
 \end{conjecture}

In any event, it is remarkable how, in spite of classes $\Cplus$ and
$\Cminus$ being at opposite extremes in the categorical classification
\eq{four}, the construction strategy utilized here has strong analogies
(blocks, layers, store-and-forward of information with repeated exchange
through block edges) with those used by Kari and Durand--Lose. This in spite
of the fact that the latter critically relies on a CA's being strictly
\emph{information lossless} while ours critically depends on their being
\emph{locally lossy}!

\Sect[consequences]{Some consequences of the conjecture}

From the truth of Conjecture \ref{conj} would derive the following remarkable
consequences.
 \begin{proposition}\label{most}
 All cellular automata, except the vanishing subset $\Czero$, can be
rewritten as isomorphic lattice gases.
 \end{proposition}

 \begin{proposition}\label{conj+}
 Regular cellular automata \emph{(see end of \sect{char})} are exactly those
which can be rewritten as isomorphic lattice gases; singular cellular
automata, those which cannot.
 \end{proposition}

\Sect[thermo]{Thermodynamic considerations}

Here we shall briefly touch on some thermodynamic aspects of the foregoing
purely mathematical results. Though we could have phrased the following
purely in terms of \emph{entropy}, without ever having to embody it into
``energy'' and ``heat,'' there will be no harm in using the more familiar,
though less general, picture.

\medskip

As mentioned right before Theorem \ref{th:lg2ca}, a concrete physical
system patterned after the ``schematics'' of a cellular automaton will
require a steady supply of power to operate and a thermal sink capable of
steadily dissipating this power---\emph{no matter whether or not the
underlying abstract dynamical system is invertible}---to support the
``document duplicating services'' continually performed by the fanout nodes
and the ``surplus document shredding'' performed by the many-input,
one-output logic gates.

Speaking here, for simplicity, as if our conjecture were true (if not, our
arguments will apply at least to the cases we have proved), for regular
cellular automata it is in principle possible to build an alternate
implementation, in the form of a lattice gas that does not require
a power supply to operate.\footnote
 {More precisely, even if some form of friction should be unavoidable because
of the nonideal behavior of the implementing mechanism, yet there is no
theoretical lower bound set by thermodynamics to the amount of this friction,
and thus to the amount of power that will have to be supplied to overcome
it.}

For those cellular automata that are invertible, besides no power supply, no
\emph{thermal sink} will be needed either, since, with Kari's and
Durand--Lose's construction, large enough lattice-gas blocks allow one to
make use of all relevant correlations and thus operate in an efficient,
dissipationless fashion. The system will work analogously to a planetary
system---a complex nonlinear system\footnote
 {Even capable of general-purpose computation, as shown in a stylized way by
Margolus's ``billiard-ball'' model of computation\cite{margolus84}.}
 that can ideally run forever on its own initial energy, without requiring an
external power source.

On the other hand, a system like Conway's ``Game of Life'' is intrinsically
noninvertible, and so to implement it within our world's physics one will
have to complement it with an ancillary ``entropy drain'' that is not
explicitly represented within the cellular-automaton model.  The combined
physical system (cellular automaton plus ancilla) is one that exhibits
friction and gradually converts ``high-grade'' energy into heat. However, our
construction shows that, much as in the invertible case, this convertion need
not entail more than the system's \emph{initial store} of energy---no
continual power injection from outside is needed for the system's
operation. Consequently, throughout the system's \emph{infinite course} of
evolution no more entropy will have to be drained from it than the
information originally contained in it at time $t=0$, which is at most $\log
a$ per site. This behavior is like that of an isolated inertial planetary
system in which some of the interactions are affected by friction.

As a dissipative system (planetary or ``Game of Life''-like) evolves and its
trajectories gradually merge, the system becomes effectively closer and
closer to being invertible\footnote
 {For instance, a Game-of-Life ``semaphore'' or a ``glider,'' taken in
isolation, may be viewed as obeying an invertible dynamics.}
 and the attendant heat production per time step decreases. The heat integral
over the entire infinite trajectory from a given initial state can never
exceed the energy contents of the initial state. In this sense, the system's
state simply ``flows downhill'' by itself, rapidly and first but increasingly
more slowly, until its mostly settles into local equilibria or local
attractor states, which are essentially nondissipative. Flickers of
dissipative nontrivial computation must still occasionally occur (if the
system is to be computation-universal), but ever more rarely and sparsely
(cf.\ Dyson\cite{dyson79}).

The perhaps surprising, but not at all paradoxical, conclusion is that a
complex CA-like dynamics can run \emph{forever}, faithfully applying its
local map at each site and each time step, fueled, as it were, just by the
negentropy of the \emph{input data}, \ie\ its own initial state.

\medskip

A physically plausible scheme for CA-like digital computation fueled by just
its input data had been proposed by Lent, Tougaw, and Porod\cite{lent93},
ostensibly realized by simple interactions between chains of bistable
one-electron quantum boxes. There was some debate at that time as to whether
such a scheme could actually work as purported---though objections mostly
reduced to ``the outcome seems too good to be true.''  The specific
interactions sketched at that time may well have been too weak to support the
desired behavior; nevertheless, our results show that a scheme like theirs is
in principle feasible, though requiring the deployment, for ``primitive
computing elements,'' of rather complex, ad hoc physical effects.

\Sect[conc]{Conclusions and perspectives}

We have presented and discussed the domain of validity of Proposition
\ref{conj+}, (a) proving that it applies to all 1D cellular automata and some
2D ones, (b) giving substantial evidence in favor of its applying to
\emph{any} number of dimensions, and (c) outlining a plausible path to
completing that evidence. In any event, just results (a) above are sufficient
to conclude that
 \begin{itemize}
 \item There exist nontrivial \emph{noninvertible} cellular automata that can
be rewritten as isomorphic lattice gases. These include cellular automata
that are computation-universal.
 \item Much as in the case of invertible cellular automata, the conversion of
a noninvertible cellular automaton into an equivalent lattice gas entails using
LG unit cells that span a possibly very large number of CA sites. The
resulting LG, then, while displaying the same \emph{functional} pitch as the
original CA, must make use of a much coarser \emph{structural} pitch---and
thus of a much more complex local mechanism.
 \item For a given computing task, the thermodynamic efficiency of a lattice
gas, which unlike a CA does not indulge in wasteful ``photocopying'' and
``shredding'' practices, can only be achieved at the cost of using LG nodes
that encompass a sufficiently large number $n$ of sites. Intuitively, such a
large node is able to match ``offer'' and ``demand'' for specific pieces of
information between faraway CA sites, routing information to successive
destinations as needed (``information recycling'') and thus eliminating
duplication and waste.
 \item The required span $n$ of the nodes of an equivalent LG (\ie\ one
capable of that efficiency) \emph{is not a computable function} of the CA's
local map $f$. Thus $n(f)$ is an \emph{unbounded} function: given an
arbitrarily large $\bar n$, there exist CA that cannot be converted into LG
with a span $n$ less than $\bar n$ sites!
 \end{itemize}

\bigskip

In \sect{back} we mentioned the use of finite-difference schemes for the
numerical integration of partial differential equations. In many cases, a
number of alternative schemes have been proposed for dealing with the
\emph{same} differential equation; of these, some may be \emph{much more
complex} than others. Why should one go to such lengths? It turns out that
the extra cost of a certain scheme may be offset by a special benefit, such
as improved convergence rate, better stability with respect to small changes
in initial conditions, or strict conservation of quantities that are
conserved in the target differential equation. Even if each ``turn of the
crank'' should more expensive, a scheme may still be competitive if many
fewer ``turns'' are needed for the same result. Or there may be schemes that
offer different tradeoffs in the space and time required for a computation
(see Bennett\cite{bennett01} and Li and Vit\'anyi\cite{li97}), and from among
which one would choose depending on the relative ``costing'' of storage and
processing resources.

In this paper we have introduced \emph{one new type of computational scheme
tradeoff}, that between the one-off cost of an efficient but complex
infrastructure and the daily drain of energy (or entropy) needed to make a
simpler, cheaper infrastructure do the same job. The spacetime regularity of
a CA's dynamics explicitly forces one to obey the golden rule, ``do unto
others as you would have them do unto you.''  That is, to improve the
thermodynamic efficiency of its own computational task, one CA site cannot
just borrow some of its right neighbor's state as a scratchpad, lest its left
neighbor claim some of \emph{his} state---a vicious infinite regress. But the
sresources of many CA sites can be pooled together into an LG ``supersite.''
Our results and conjecture deal with how, for any regular CA, a large enough
LG supersite can be constructed that achieves \emph{perfect thermodynamic
efficiency}. To paraphrase \cite{economist89}, we may term this approach
``logic for capitalists.''

\medskip

Having dispensed with a \emph{power supply} for locally lossy CA, if the
resulting LG system has to be embedded within an invertible physics one must
still complement it with a \emph{heat sink}. Our construction leaves open the
issue of how to realize such a sink while retaining the local finiteness of
an LG scheme. Can removing and sequestering away a finite amount of
heat per site be achieved by a finite amount (per site) of extra storage and
processing resources by an LG mechanism---and thus by \emph{local} means?
This is a wide open area for research.

\medskip

Finally, there is a rather delicate question that we can barely mention here.
We recall that computation is the art of putting together---even though in as
large a number of copies as desired---a finite set of logic primitives
\emph{given once and for all.}  \emph{Within} any LG this condition is
clearly satisfied, as in an LG all processing is done by composition of
instances of a \emph{single} unit cell, complex as this may be. But to what
extent can one accept as \emph{bona fide computation} an approach that
requires designing, for each CA that one wants to rewrite as an LG, new
custom primitives, that is, the nodes that make up the LG's very unit cell?
Note that these may be arbitrary $n$-input, $n$-output boolean functions.

Among the physical ``effects'' actually available in nature one may well hope
to find one that can be exploited for realizing a simple logic primitive such
as the {\sc and} gate (or an invertible counterpart thereof, such as the so
called ``Toffoli gate'' \cite{toffoli80rc,barenco95}). But to what extent
will we be able to locate in nature ``custom Hamiltonians'' capable of
realizing (without power assist) arbitrarily complex $n$-body interactions?
Intuitively, information recycling can make computation substantially more
efficient (cf.\ \cite{economist89}). But, at least within the stylized
context of CA and LG, we've touched with hand reasons why recycling schemes
may become impractical well before one attains perfect thermodynamic
efficiency.

\section*{Acknowledgments}

One of us (T. Toffoli) would like to thank the Mathematics Institute of the
University of Rome ``La Sapienza'' (Italy) for a visiting professorship that
provided the occasion for this research; acknowledge the stimulus received
from Charles H. Bennett, Ed Fredkin, Lev Levitin, Norman Margolus, and
Michael Frank in the pursuit of this kind of investigation; and acknowledge
the participation of Alex Ashpunt, Tony Short, and Sandu Popescu in related
work in progress.

 \end{document}

 {\rm (Fekete)}\quad Let $g:\natural\to\natural$ be a function for which
$g(i+j)\leq g(i)+g(j)$ for all $ij\in\natural$. Then
$\lim_{i\to\infty}g(i)/i$ exists.

, and
$\kappa(i)=1-v(i)/i$ its variety defect \emph{per site}

(where \emph{decidable} is synonymous with `recursive' and
\emph{semidecidable} with `recursively enumerable')